\documentclass{amsart}
\usepackage{amssymb,epsfig,amsthm}
\usepackage{times}
\newtheorem{lemma}{Lemma}[section]
\newtheorem{theorem}[lemma]{Theorem}

\begin{document}

\title[Charged dust disks]{On explicit solutions to the stationary axisymmetric 
Einstein-Maxwell equations describing dust disks
}
\author[C.~Klein]{C.~Klein\address{
     Max-Planck-Institut f\"ur Physik, F\"ohringer Ring 
    6, 80805 M\"unchen, Germany}        
}
\date{\today}    

\begin{abstract}
    We review explicit solutions to the stationary axisymmetric
    Einstein-Maxwell equations 
    which can be interpreted as disks of charged dust. The disks 
    of finite or 
    infinite extension are infinitesimally thin and 
    constitute a surface layer at the boundary of an electro-vacuum. 
    The Einstein-Maxwell equations in the presence of one Killing 
    vector are obtained by using a 
    projection formalism. 
    This leads to equations for three-dimensional gravity 
    where the matter is given by a $SU(2,1)/S[U(1,1)\times U(1)]$ 
    nonlinear sigma model. The $SU(2,1)$ invariance of the stationary 
    Einstein-Maxwell equations can be used to construct solutions for 
    the electro-vacuum from solutions to the pure vacuum case via a 
    so-called Harrison transformation. It is 
    shown that the corresponding solutions will always have a 
    non-vanishing total charge and a gyromagnetic ratio of 2. 
    Since the vacuum and the electro-vacuum equations in the stationary 
    axisymmetric case are completely integrable, large 
    classes of solutions can be constructed with techniques from the 
    theory of solitons. The richest class of physically interesting 
    solutions to the pure vacuum case due to Korotkin is given in 
    terms of hyperelliptic theta functions. The Harrison transformed 
    hyperelliptic solutions are discussed. As a concrete example we 
    study the transformation of a family 
    of counter-rotating dust disks. To obtain 
    algebro-geometric solutions with vanishing total 
    charge which are of astrophysical relevance, three-sheeted 
    surfaces have to be considered. 
    
    The matter in the disk is 
    discussed following Bi\v{c}\'ak et al. We review the 
    `cut and glue' technique where a strip is removed from an 
    explicitly known spacetime and where the remainder is 
    glued together after displacement. The discontinuities of the 
    normal derivatives of the metric at the glueing
    hypersurface lead to infinite disks. If the 
    energy conditions are satisfied and if the pressure is positive, 
    the disks can be interpreted in 
    the vacuum case as made up of two components of counter-rotating 
    dust moving on geodesics. In electro-vacuum the condition 
    of geodesic movement is replaced by electro-geodesic movement. 
    As an 
    example we discuss a class of Harrison-transformed hyperelliptic 
    solutions. The range of parameters is identified
    where an interpretation of the 
    matter in the disk in terms of electro-dust can be given. 
\end{abstract}

\maketitle

\section{Introduction}
Electromagnetic fields, especially magnetic fields play a role in 
astrophysics in the context of neutron stars, white dwarfs 
and galaxy formation. A complete relativistic understanding of 
such situations requires studying the coupled 
Einstein-Maxwell equations. Of special interest are stationary 
axisymmetric situations since isolated matter configurations in 
thermodynamical equilibrium belong within relativity to this class, 
see \cite{hartle,lindblom}. 
Since the stationary axisymmetric Einstein-Maxwell equations in vacuum 
in the form of Ernst \cite{ernst2}
are completely integrable (see \cite{maison79}), powerful solution 
generating techniques from the theory of solitons 
are at hand to obtain physically interesting 
solutions. But the equations in the matter region --- which is 
generally approximated as an ideal fluid ---  do not seem to be 
integrable. This makes it difficult to find 
global solutions which hold 
both in a three-dimensionally extended matter region  and in vacuum.

To obtain global solutions to the Einstein-Maxwell equations with 
these methods, one is thus limited to two-dimensionally extended 
matter distributions, i.e.\ surface layers. Infinitesimally thin 
disks have been discussed in Newtonian astrophysics as models for 
certain galaxies, see \cite{bitr},  and as models for the 
matter in accretion disks around black holes, see \cite{semerak} and 
references therein. In this case the equations in the 
matter region reduce to ordinary differential equations the solutions 
of which determine boundary data for the vacuum equations.  
Alternatively surface layers can be obtained by `cut and glue' 
techniques. Bi\v{c}\'ak, 
Lynden-Bell and Katz \cite{blk93,blk} studied static spacetimes, from 
which they removed a strip and glued the remainder together. 
The non-continuous normal derivatives at the gluing
plane lead to a $\delta$-type energy-momentum tensor which can be 
interpreted as an infinitely extended disk made up of 
counter-rotating dust. This method was extended in \cite{ledvinka} to 
generate disk sources of the Kerr-metric. With the same techniques, 
disk sources for Kerr-Newman metrics \cite{zofka}, static axisymmetric 
spacetimes with magnetic fields \cite{letelier} and conformastationary 
metrics \cite{kbl99} were given.  

Counter-rotating disks 
are discussed in astrophysics 
as models for certain $S0$ and $Sa$ galaxies (see 
\cite{galaxies} and references given therein). These galaxies show 
counter-rotating matter components and are believed to be the 
consequence of the merger of galaxies. Recent investigations have 
shown that there is a large number of galaxies (see \cite{galaxies}, 
the first was NGC 4550 in Virgo) which show counter-rotating streams 
in the disk with up to 50 \% 
counter-rotation.

By construction all disks due to `cut and glue' techniques have 
an infinite extension but finite mass since the mass of the spacetime 
is not changed by the method. The matter in the disks can be 
interpreted as a two-dimensional fluid with a purely azimuthal pressure. 
If the energy conditions in the disk are satisfied and if the 
pressure in the disk is positive, the matter can 
alternatively be interpreted as consisting of two counter-rotating 
streams of pressureless matter, so-called dust. In the pure vacuum 
case this is best done by introducing observers rotating with the 
disk in a way that the energy-momentum tensor is diagonal for them. It 
can be shown that the corresponding dust streams move on geodesics of 
the inner geometry of the disk. In the electro-vacuum case, the 
corresponding condition on the matter is motion on 
electro-geodesics, i.e.\ solutions to the geodesic equation in the 
presence of a Lorentz force. It was shown in \cite{zofka} that this 
is a more restrictive condition than in the pure vacuum case. 

For vacuum so-called Riemann-Hilbert 
techniques (see \cite{koro88,prd,jgp2} and references therein) were used to 
generate solutions for disks of finite extension. Explicit metrics 
could be given in terms of  theta functions on hyperelliptic 
Riemann surfaces \cite{koro88}. Since these two-sheeted surfaces are a generalization of 
the well-known elliptic surfaces, a powerful theory is at hand to 
treat hyperelliptic functions. The main advantages of this class are 
that it is very rich (`solitonic' solutions as the Kerr solution 
for a rotating black hole are contained as limiting cases), 
and that a whole subclass 
with physically interesting solutions could be identified in 
\cite{prl,prd2} by studying the analyticity properties of the solution. 
The corresponding solutions for the 
electro-vacuum are given on three-sheeted surfaces \cite{koro88} which 
are mathematically less well understood. Recent progress in this 
context was made by Korotkin \cite{koro03} 
by considering the Riemann-Hilbert problem on multi-sheeted coverings 
of the complex plane. The solutions to this problem which can be 
used to solve the Einstein-Maxwell equations are again given in 
terms of theta functions. 

Until now there are, however, no explicit examples for 
physically realistic disk solutions on three-sheeted Riemann surfaces. 
Therefore an intermediate step was taken in \cite{prharrison} where 
hyperelliptic solutions with charge were studied. These solutions were 
obtained by exploiting the $SU(2,1)$ invariance of the stationary 
Einstein-Maxwell equations 
(see \cite{nk} to \cite{bgm}). Using a so-called Harrison 
transformation \cite{harrison}, 
one can generate solutions with charge from solutions 
to the pure vacuum equations.  A remarkable property of 
the corresponding 
spacetimes is the fact that their gyromagnetic 
ratio is always identical to 2 as in the case of the 
Kerr-Newman black holes. This is of interest in the context 
of claims in \cite{kipfi02,kipfi03} that this property
is a hint on a deep 
connection between general relativity and relativistic quantum 
mechanics. By studying the asymptotics of Harrison-transformed 
pure vacuum solutions, 
it was shown that the thus obtained solutions will always 
have a non-vanishing total charge which limits their astrophysical 
relevance since charges seem to neutralize in our universe. 
This is a hint that 
astrophysically interesting solutions without total charge, but 
non-vanishing magnetic fields in terms of theta functions are  
to be expected only on three-sheeted surfaces.

This paper is is organized as follows: In section 2 the Newtonian case 
is studied for illustration. The `cut and glue' techniques of 
Bi\v{c}\'ak et al.\ \cite{blk93} and disks of finite extension
are presented. In section 3 we consider the 
Einstein-Maxwell equations in the presence of a single Killing 
vector. Using a projection formalism \cite{ehlers,geroch}, 
we perform a standard 
dimensional reduction of the equations. It is shown that the 
stationary Einstein-Maxwell equations are equivalent to
three-dimensional gravity with a $SU(2,1)/S[U(1,1)\times U(1)]$ sigma 
model as matter. Using complex notation, one can introduce Ernst 
potentials \cite{ernst2}. We study a gauge invariant formulation of 
the $SU(2,1)$ matrix of the sigma model
and the related transformations of the 
solutions. The Harrison transformation is presented and discussed for 
simple examples. The asymptotic behavior of asymptotically flat solutions is 
studied. The stationary axisymmetric case is shown to be completely 
integrable. In section 4 we recall basic facts on the stationary 
axisymmetric pure vacuum case and on hyperelliptic disk
solutions. Using the Harrison transformation 
on a family of counter-rotating disk solutions \cite{prl2,prd3}, we 
obtain the complete transformed 
metric and discuss interesting limiting cases. The 
discussion of the energy-momentum tensor using Israel's junction 
conditions \cite{israel} is presented in section 5. 
The case of the Harrison transformed counter-rotating disk is 
studied as an example. The range of the physical parameters
is given where 
the matter in the disk can be interpreted as electro-dust.
In section 6 we summarize recent results by Korotkin \cite{koro03} on 
solutions to the
Riemann-Hilbert problem on multi-sheeted Riemann surfaces in terms of 
Szeg\"o kernels and solutions to the Einstein-Maxwell equations. 
In section 7 we add some concluding remarks.

\section{Newtonian dust disks}\label{sec.2}
To illustrate the basic concepts used in the following sections, we 
will briefly recall some facts on Newtonian dust disks. In 
Newtonian theory, gravitation is described by a scalar potential $U$ 
which is a solution to the Laplace equation in the vacuum region. 
The units in this article are chosen in a way that the 
Newtonian gravitational constant, the dielectric constant 
and the velocity of light are equal 
to 1.
We use cylindrical coordinates $\rho$, $\zeta$ 
and $\phi$ and place the disk made up of a pressureless 
two-dimensional ideal fluid with radius $\rho_{0}$ in the equatorial 
plane $\zeta=0$.  In 
Newtonian theory stationary perfect fluid solutions and thus also the 
here considered disks are known to be equatorially symmetric.

Since we concentrate on dust disks, i.e.\ pressureless 
matter, the only force to compensate gravitational attraction 
in the disk is the centrifugal force. This leads in the disk to (here 
and in the following $f_{,x}=\frac{\partial f}{\partial x}$)
\begin{equation}
	U_{,\rho}=\Omega^{2}(\rho)\rho,
	\label{eq1}
\end{equation}
where $\Omega(\rho)$ is the angular velocity of the dust at radius 
$\rho$. Since all terms in (\ref{eq1}) are quadratic in $\Omega$ 
there are no effects due to the sign of the angular velocity. The 
absence of  
these so-called gravitomagnetic effects in Newtonian 
theory implies that disks with counter-rotating components 
will behave with respect to gravity exactly as 
disks made up of only one component. We will therefore only 
consider the case of one component in this section.
Integrating (\ref{eq1}) we get the boundary data $U(\rho,0)$ with an
integration constant $U_{0}=U(0,0)$ which is related to the central redshift in 
the relativistic case. 

To find the Newtonian solution for a given rotation law 
$\Omega(\rho)$, we 
have to construct a solution to the Laplace equation which is 
everywhere regular except at the disk where it has to take the 
boundary data (\ref{eq1}). At the disk the normal 
derivatives of the potential will have a jump since the disk is a 
surface layer. Notice that one only has to solve the vacuum equations 
since the two-dimensional matter distribution merely leads to boundary 
conditions for the Laplace equation. In the Newtonian setting one thus has 
to determine the density for a given rotation law or vice versa, a 
well known problem (see e.g.\ \cite{bitr} and references therein) 
for Newtonian dust disks.

There are several ways to construct Newtonian dust disks. We will only 
outline two possibilities which can be used with some modifications 
also in the relativistic case. 

\subsection{`Cut and glue'-techniques}
One way to construct Newtonian disks is to start with a known 
equatorially symmetric solution to the Laplace equation, 
for instance the solution for a point mass $m$,
\begin{equation}
    U= -\frac{m}{\sqrt{\rho^{2}+\zeta^{2}}}.
    \label{eq:point}
\end{equation}
Then a strip of width $2\zeta_{0}$ 
is cut out of the space symmetrically to 
the equatorial plane. The solutions for positive and negative $\zeta$ 
are displace in $\zeta$-direction by $\pm \zeta_{0}$ and glued 
together at the equatorial plane. The discontinuity at this plane 
leads to a surface layer of infinite extension.

In the example of a point mass considered by Kuzmin \cite{kuzmin} 
and Toomre \cite{toomre}, this leads to the solution 
\begin{equation}
    U=-\frac{m}{\sqrt{\rho^{2}+(|\zeta|+\zeta_{0})^{2}}}
    \label{eq:point2}.
\end{equation}
The surface density $\sigma_{d}$ at the equatorial plane is just given 
by $2\pi\sigma_{d}=U_{\zeta}(0^{+})$, in the example
\begin{equation}
    U_{\zeta}(0^{+})=\frac{m\zeta_{0}}{
    (\rho^{2}+\zeta_{0}^{2})^{\frac{3}{2}}}
    \label{eq:point3}.
\end{equation}
By construction the spacetime has finite mass $m$ since the 
asymptotics have not been changed by the procedure of cutting and 
glueing. The disk is infinite, but the mass density decreases because 
of (\ref{eq:point3}) as $\rho^{-3}$. Since the mass density is 
positive, the matter in the disk can be interpreted as a disk of dust. 
The angular velocity as defined in (\ref{eq1}) is in this example given  
by 
\begin{equation}
    \Omega^{2}=\frac{m}{(\rho^{2}+\zeta_{0}^{2})^{\frac{3}{2}}}
    \label{eq:point4}.
\end{equation}
Asymptotically the angular velocity satisfies the Kepler relation 
$\Omega^{2}\rho^{3}=m$ for test particles. Thus the matter in the disk 
behaves  for large distances from the center as free particles. Though 
the gravitational field (\ref{eq:point}) can be seen as generated by 
the self-gravitating matter in the disk, particles in large distances 
from the center can be viewed as test particles since the the density 
tends to zero. 
The so constructed disks thus have physically acceptable properties: a 
positive mass density and a finite mass. Though they have an infinite 
extension, the density decreases rapidly for $\rho\to\infty$. 

Since the Laplace equation is linear, arbitrary linear combinations 
of potentials of the form (\ref{eq:point}) will always leads to 
infinite disks.  Evans and de Zeeuw \cite{evazee} 
considered disk potentials of the form 
\begin{equation}
    U = \int\frac{\nu(\epsilon)d\epsilon}{\sqrt{\rho^{2}+
    (|\zeta|+\epsilon)^{2}}}
    \label{eq:point5}
\end{equation}
which leads to the general classical disk formula.  Bi\v{c}\'ak, 
Lynden-Bell and Katz \cite{blk93} used this technique to generate 
disk solutions to the static axisymmetric Einstein equations.

\subsection{Disks of finite extension}
To generate disks of finite extension, we use an approach which can 
be generalized to some extent to the relativistic case. The resulting 
expression will be shown to be equivalent to the Poisson integral for 
a distributional density. We put 
$\rho_{0}=1$ without loss of generality (we are only considering disks 
of finite non-zero radius) and obtain $U$ as the solution of a 
so-called 
Riemann-Hilbert problem (see e.g.\ \cite{jgp2} and references given 
therein). The solution can be written in the form
\begin{equation}
	U(\rho,\zeta)=-\frac{1}{4\pi 
	\mathrm{ i}}\int_{\Gamma}^{}\frac{\ln G(K) dK}{\sqrt{(K-\zeta)^2+
	\rho^2}}
	\label{newton2},
\end{equation}
where
$\ln G\in C^{1,\alpha}(\Gamma)$ (H\"older continuous on $\Gamma$) 
and where $\Gamma$ is the covering of the 
imaginary axis in the upper sheet of $\mathcal{L}$ 
between $-\mathrm{ i}$ 
and $\mathrm{ i}$;  $\mathcal{L}$ is the Riemann surface of genus 0 
given 
by the algebraic relation $\mu_{0}^{2}(K)=(K-\zeta)^2+\rho^{2}$. 
The function $G$ has to be subject to the conditions 
$G(\bar{K})=\bar{G}(K)$ and
$G(-K)=G(K)$. 

It may be checked by direct calculation that $U$ in (\ref{newton2}) 
is a solution to the Laplace equation except at the disk. The 
reality condition on $G$ 
leads to a real potential, whereas the symmetry condition with 
respect to the involution $K\to -K$ leads to equatorial 
symmetry.  
The occurrence of the logarithm in (\ref{newton2}) is due to the 
Riemann-Hilbert problem with the help of which the solution to the Laplace 
equation was constructed, see e.g.\ \cite{jgp2}.

The function $\ln G$ is determined by the boundary data 
$U(\rho,0)$ or the energy density $\sigma_{d}$ of the dust 
 via
\begin{equation}
	\ln G(t) = 4\left(U_0+t\int_{0}^{t}
		\frac{U_{\rho}(\rho)d\rho}{\sqrt{t^2-\rho^2}}\right)
	\label{eq4}
\end{equation}
or
\begin{equation}
	\ln G(t)=4 \int_{t}^{1}\frac{\rho U_{\zeta}}{\sqrt{\rho^{2}-t^{2}}}d\rho
	\label{eq5}
\end{equation}
respectively where $t=-iK$. This can be seen in the 
following way:\\
At the disk the potential takes due to the equatorial symmetry
the boundary values
\begin{equation}
	U(\rho,0)=-\frac{1}{2\pi }\int_{0}^{\rho}
	\frac{\ln G(t) }{\sqrt{\rho^2-t^{2}}}dt
	\label{newton3}
\end{equation}
and 
\begin{equation}
	U_{\zeta}(\rho,0)=-\frac{1}{2\pi 
	}\int_{\rho}^{1}\frac{\partial_{t}(\ln G(t))}{\sqrt{t^{2}-\rho^{2}}}dt
	\label{eq2}.
\end{equation}
Both equations 
constitute integral equations for the `jump data' $\ln G$ of the 
Riemann-Hilbert problem if the respective left-hand side is known. 
The equations (\ref{newton3}) and (\ref{eq2})
are both Abelian integral equations and can be solved in 
terms of quadratures, i.e.\  (\ref{eq4}) and (\ref{eq5}).

To show the regularity of the potential $U$, we prove that the 
integral (\ref{newton2}) is  identical to the 
Poisson integral for a distributional density which reads at the disk
\begin{equation}
	U(\rho)=-2\int_{0}^{1}\sigma_{d}(\rho')\rho' d\rho' \int_{0}^{2\pi}
	\frac{d\phi}{\sqrt{(\rho+\rho')^{2}-4\rho\rho' \cos \phi}}
	=-4 \int_{0}^{1}\sigma_{d}(\rho')\rho' d\rho'\frac{K(k(\rho,\rho'))}{\rho+\rho'},
	\label{eq7}
\end{equation}
where $k(\rho,\rho')=2\sqrt{\rho\rho'}/(\rho+\rho')$ and where $K(k)$ is the 
complete elliptic integral of the first kind.  Eliminating $\ln G$ in 
(\ref{newton3}) via (\ref{eq5}) we obtain after interchange of the 
order of integration
\begin{equation}
	U=-\frac{2}{\pi}\left(\int_{0}^{\rho}U_{\zeta}\frac{\rho'}{\rho}
	K\left(\frac{\rho'}{\rho}\right)d\rho'+\int_{\rho}^{1}U_{\zeta}
	K\left(\frac{\rho}{\rho'}\right)d\rho'\right)
	\label{eq8}
\end{equation}
which is identical to (\ref{eq7}) since $K(2\sqrt{k}/(1+k))=(1+k)K(k)$. 
Thus the integral (\ref{newton2}) has the 
properties known from the Poisson integral: it is 
a solution to the Laplace equation which is everywhere 
regular except at the disk where the normal derivatives are 
discontinuous.

We note that it is possible in the Newtonian case
to solve the boundary value problem purely locally at the disk. 
The regularity properties of the Poisson integral then ensure global 
regularity of the solution except at the disk. Such a purely local 
treatment will not be possible in the relativistic case.

The above considerations make clear that one cannot prescribe 
both $U$ at the disk (and thus the rotation law) and the density 
independently. This just reflects the fact that the Laplace equation 
is an elliptic equation for which Cauchy problems are ill-posed.
If $\ln G$ is determined by either (\ref{eq4}) or (\ref{eq5}) for 
given rotation law or density, expression 
(\ref{newton2}) gives the analytic continuation of the boundary data 
to the whole spacetime. In case we prescribe the angular velocity, the 
constant $U_{0}$ is determined by the condition $\ln G(i)=0$ which 
excludes a ring singularity at the rim of the disk. For rigid 
rotation ($\Omega=const$), we get e.g.\
\begin{equation}
	\ln G(\tau)=4\Omega^2(\tau^2+1),
	\label{eq6}
\end{equation}
which leads with (\ref{newton2}) to the well-known Maclaurin disk.

\subsection{Charged static dust disks}
In a Newtonian theory gravity and electromagnetism decouple. 
Disks of pressureless charged matter will lead to a gravitational 
potential $U$ as above and to electric and magnetic fields. For disks 
with only one component of dust, there will be necessarily a magnetic 
field due to the rotating charges. Static solutions are possible if 
two streams of charged dust with equal densities are exactly 
counter-rotating. In this case the magnetic field vanishes since the 
effects of both streams concerning magnetic fields just compensate. 
This trick was used by Morgan and Morgan \cite{morgan} 
within general relativity to describe static disks: in the case of
two identical 
streams of counter-rotating particles the so-called 
`gravitomagnetic' effects of relativity cancel, and 
the resulting spacetime is static. 

In the electrostatic case, the electric potential is also a solution 
to the Laplace equation. 
If we assume the electrical density $\sigma_{e}$ 
to be proportional to the matter density $\sigma_{d}$ in the case of a dust 
disk (both densities are surface densities), i.e.\ $\sigma_{e}=
Q\sigma_{d}$, we 
get from Newton's law $F_{grav}=F_{centrifugal}+F_{el}$ the relation
\begin{equation}
    U_{\rho}(1-Q^{2})=\Omega^{2}\rho
    \label{newton}.
\end{equation}
Thus one can infer the tangential derivative at the disk for given 
$\Omega$ and constant $Q$ and then solve the boundary value problem 
for the Laplace equation as above. The electric and the gravitational 
potential are in this example just proportional.  
Similarly one can prescribe 
$\sigma_{d}$ and obtain $\Omega$ from (\ref{newton}). Note that $Q^{2}$ 
has to be smaller than 1. For $Q^{2}=1$, the angular velocity in the 
dust disk vanishes.

\section{Einstein-Maxwell equations and group structure}
In this section we study the Einstein-Maxwell equations in the 
presence of one Killing vector. We explore the group structure of the 
equations and give the Harrison transformation which generates 
electro-vacuum solutions from pure vacuum solutions. The solutions 
contain an additional real parameter related to the total charge. General 
properties of the transformed spacetimes as the asymptotics are 
discussed. In the stationary axisymmetric case, complete integrability 
of the equations is established.
\subsection{Maxwell equations}
It is instructive to consider first the Maxwell equations in the 
absence of gravity, i.e.\ on a flat background. In standard notation, 
the equations for the electric and respectively magnetic fields $\mathbf{E}$ 
and $\mathbf{B}$ read
\begin{equation}
    \mbox{div} \mathbf{E}=0, \quad \mbox{rot}\mathbf{E}+
    \mathbf{B}_{,t}=0, \quad \mbox{div}\mathbf{B}=0,
    \quad \mbox{rot}\mathbf{B}-\mathbf{E}_{,t}=0.
    \label{eq:maxwell}
\end{equation}
Since $\mbox{div}\mathbf{B}=0$, one can define (up to gauge freedoms) 
a vector potential via $\mathbf{B}=\mbox{rot}\mathbf{A}$. 
It is convenient for the relativistic treatment in the following 
sections to introduce four-dimensional notation. We 
use the convention that 
greek indices take the values $0,1,2,3$ and latin indices 
the values $1,2,3$.
A four-dimensional vector potential $A_{\mu}=(A,A_{a})$ 
is introduced which is related to the tensor 
$F_{\mu\nu}$ of the electromagnetic fields via 
\begin{equation}
    F_{\mu\nu}=A_{\mu,\nu}-A_{\nu,\mu}.
    \label{eq:fa}
\end{equation}
In vacuum, the Maxwell equations read (indices are raised and lowered 
with the Minkowski metric)
\begin{equation}
    F_{\mu\nu}{}^{,\nu}=0, \quad ^{*}F_{\mu\nu}{}^{,\nu}=\frac{1}{2}
    \epsilon_{\mu\nu\alpha\beta}F^{\alpha\beta}{}^{,\nu}=0
    \label{maxflat}.
\end{equation}
They can be obtained from the action
\begin{equation}
    S = \frac{1}{4}
    \int_{}^{}dx^{4}F_{\mu\nu}
    F^{\mu\nu}
    \label{eq:action0}.
\end{equation}

Obviously the equations (\ref{max}) 
are invariant under the discrete exchange 
$F\to {}^{*}F$. In addition they are invariant under the continuous 
rotations 
\begin{equation}
    F+i{}^{*}F\to e^{i\theta}(F+i{}^{*}F), \quad \theta\in \mathbb{R}
    \label{eq:u1},
\end{equation}
or in terms of the fields, $\mathbf{E}+i\mathbf{B}\to e^{i\theta}
(\mathbf{E}+i\mathbf{B})$. This is the 
well known $U(1)$ symmetry of the Maxwell equations in vacuum. It can 
be used to `generate' solutions from known ones: if for instance 
a solution with vanishing magnetic field is given, the 
$U(1)$-symmetry can be applied to generate from the given electric field 
a solution with non-vanishing magnetic field. The transformed fields 
contain a new real parameter.  In case there is an 
electric monopole moment, there will be a magnetic 
monopole moment in the transformed solution. This often limits the
physical relevance of the so generated solutions. 

In the stationary case, the potential $A_{\mu}$ can be chosen to be 
independent of the time coordinate $t$. Since 
$\mbox{rot}\mathbf{B}=0$, we can define a scalar potential via 
$\mathbf{B}=\mbox{grad}B$, the well-known magnetic potential 
for stationary fields. The equation $\mbox{div}\mathbf{B}=0$ 
implies $\Delta B=0$. In four-dimensional language which will be 
needed in the Einstein-Maxwell case, this construction works as 
follows: Since $F^{ab}$ is a 
three-dimensional antisymmetric tensor, it can be dualized to an 
axial vector by contraction with the totally antisymmetric 
$\epsilon$-tensor. The Maxwell equations 
(\ref{max})  $F^{ab}{}_{,b}=0$ imply that this vector must be a 
gradient of some potential $B$. 
We can thus define the potential $B$ via
\begin{equation}
    B_{,c}=-\epsilon_{abc}A_{a,b}
    \label{eq:B}.
\end{equation}
The 
potentials can be combined to the complex potential $\Phi=A+iB$. In 
this case the action is just given by 
\begin{equation}
    S = \frac{1}{2}
    \int_{}^{}dx^{3}|\nabla\Phi|^{2}
    \label{eq:action0a},
\end{equation}
and the Maxwell equations read 
\begin{equation}
    \Delta \Phi=0
    \label{eq:max2}.
\end{equation}
In the stationary case, the Maxwell 
equations are thus equivalent to the 
Laplace equation for a single complex potential $\Phi$.

\subsection{Einstein-Maxwell equations}
In the Einstein-Maxwell case, the Maxwell equations have the same 
form as in (\ref{max}), only the partial derivatives have to be 
replaced by covariant derivatives since the spacetime is no longer 
flat,
\begin{equation}
    F_{\mu\nu}{}^{;\nu}=0, \quad ^{*}F_{\mu\nu}{}^{;\nu}=0
    \label{max}.
\end{equation}
The tracefree energy-momentum tensor of the electromagnetic 
field is given by 
\begin{equation}
    T_{\mu\nu}=F_{\mu\alpha}F_{\nu}{}^{\alpha}-\frac{1}{4}g_{\mu\nu} 
F_{\alpha\beta}F^{\alpha\beta}
    \label{eq:enmom}.
\end{equation}
The 
Einstein equations have the form 
\begin{equation}
    R_{\mu\nu}-\frac{1}{2}g_{\mu\nu}R=T_{\mu\nu}
    \label{eq:einstein},
\end{equation}
where $g_{\mu\nu}$ is the metric of the spacetime, $R_{\mu\nu}$ the Ricci 
tensor and $R$ the Ricci scalar. 
With (\ref{eq:enmom}) we get for (\ref{eq:einstein})
\begin{equation}
    R_{\mu\nu}=F_{\mu\lambda}F^{\lambda}_{\nu}-\frac{1}{4}g_{\mu\nu}
    F_{\kappa\lambda}F^{\kappa\lambda}
    \label{einstein}.
\end{equation}
Equations (\ref{max}) and (\ref{einstein}) form the Einstein-Maxwell 
equations. They can be derived from the action
\begin{equation}
    S = \frac{1}{2}
    \int_{}^{}dx^{4}\sqrt{-g}\left(R-\frac{1}{2}F_{\mu\nu}
    F^{\mu\nu}\right),\quad g=\det (g_{\mu\nu})
    \label{eq:action1}.
\end{equation}
Since the Maxwell fields only enter the Einstein equations via the 
energy-momentum tensor, the $U(1)$ symmetry (\ref{eq:u1}) carries 
over to the Einstein-Maxwell case and can be used to generate 
solutions as before.

In general one would expect that the above $U(1)$ invariance of the 
Einstein-Maxwell equations is the only symmetry of the equations even 
in the presence of Killing symmetries in the spacetime. However, it 
turns out that a much bigger symmetry group exists already 
for a 
single Killing vector. It is convenient to use a projection formalism 
which goes back to Ehlers \cite{ehlers}, see also \cite{geroch} and 
\cite{maisonehlers} for additional references. The case considered 
here can be viewed as a special case of the standard dimensional 
reduction of Kaluza-Klein \cite{maisonehlers} 
and supergravity theories \cite{supergravity}. 
In this formalism, the metric is written in the form
\begin{equation}
    ds^{2}=-f(dt+k_{a}dx^{a})(dt+k_{b}dx^{b})+\frac{1}{f}h_{ab}dx^{a}
    dx^{b}
    \label{maison1}.
\end{equation}
We are considering here for convenience the stationary case, i.e.\ we 
use coordinates in which the timelike Killing vector is given by 
$\partial_{t}$, all potentials are independent of $t$.
However, the results hold with minor changes 
for general Killing vectors. Coordinates are 
introduced for convenience, the dimensional reduction can also 
be carried out in a coordinate-independent way. We assume that the 
Killing vector is timelike throughout the spacetime, i.e.\ that its 
norm  $f$ does not vanish.

The vector potential is decomposed as the metric into pieces parallel 
and orthogonal to the Killing vector, 
$A_{\mu}=(A,\mathcal{A}_{m}+k_{m}A)$. 
The Lagrangian of (\ref{eq:action1}) can then be written in the form 
\begin{equation}
    \mathcal{L}=\frac{1}{2}\sqrt{h}\left(\mathcal{R}-
    \frac{1}{2f^{2}}h^{ab} 
    f_{,a}f_{,b}+ \frac{f^{2}}{4}\mathcal{K}_{ab}\mathcal{K}^{ab} +
    \frac{1}{f}h^{ab}A_{,a}A_{,b}-\frac{f}{2}(\mathcal{F}_{ab}+A
    \mathcal{K}_{ab})(\mathcal{F}^{ab}+A\mathcal{K}^{ab})\right)
    \label{eq:langrangean}
\end{equation}
where $\mathcal{L}$ is a three-dimensional Lagrangian density, 
where $\mathcal{F}_{ab}=\mathcal{A}_{a,b}-\mathcal{A}_{b,a}$, and 
where $\mathcal{K}_{ab}=k_{a,b}-k_{b,a}$. All indices are raised and 
lowered with $h_{ab}$. 
Note that the tensor $\mathcal{K}_{ab}$ vanishes only if the Killing 
vector is hypersurface orthogonal in which case the spacetime is 
static. 

The first part of the Maxwell 
equations (\ref{max}) can be written in the form 
\begin{equation}
    \frac{1}{\sqrt{-g}}(\sqrt{-g}F^{\mu\nu})_{,\nu}=0
    \label{eq:maxpartial}
\end{equation}
which implies $(\sqrt{h}F^{ab}/f)_{,b}=0$. With this relation or by 
varying (\ref{eq:langrangean}) with respect to $\mathcal{A}_{a}$,
we obtain
\begin{equation}
    (\sqrt{h}f(\mathcal{F}^{ab}+A\mathcal{K}^{ab}))_{,b}=0.
    \label{eq:maxspace}
\end{equation}
We can define as before the potential $B$ via 
\begin{equation}
    B_{,c}=-\frac{1}{2}\epsilon_{cab}\sqrt{h}f(\mathcal{F}^{ab}+A\mathcal{K}^{ab})
    \label{eq:B2}.
\end{equation}
Again $A$ and $B$ can be combined to the complex electromagnetic 
potential $\Phi=A+iB$. 

Similarly we get by varying (\ref{eq:langrangean}) with respect to 
$k_{a}$
\begin{equation}
    \left(\sqrt{h}\left(\frac{f^{2}}{2}\mathcal{K}^{ab}-Af
    (\mathcal{F}^{ab}+A\mathcal{K}^{ab})\right)\right)_{,b}=0
    \label{eq:Kab}.
\end{equation}
This can be dualized as above by introducing the so-called twist 
potential $b$ via
\begin{equation}
    b_{,c}=\epsilon_{cab}\sqrt{h}\frac{f^{2}}{2}\mathcal{K}^{ab}
    +BA_{,c}-AB_{,c}
    \label{eq:twist}.
\end{equation}
The potentials $f$ and $b$ can be combined to the 
complex Ernst potential,
\begin{equation}
    \mathcal{E}=f-\Phi\bar{\Phi}+ib
    \label{eq:ernstpot}.
\end{equation}

The scalars $b$ and $B$ replace the vectors $k_{a}$ and 
$\mathcal{A}_{a}$. The corresponding three-dimensional Lagrangian 
reads with $w_{a}=b_{,a}-2BA_{,a}+2AB_{,a}$
\begin{equation}
    \mathcal{L}=\frac{\sqrt{h}}{2}\left(\mathcal{R}-h^{ab}\left(
    \frac{1}{2f^{2}}(f_{,a}f_{,b}+w_{a}w_{b})-\frac{1}{f}
    (A_{,a}A_{,b}+B_{,a}B_{,b})\right)\right)
    \label{eq:actint}.
\end{equation}
The line element 
\begin{equation}
    ds^{2}=\frac{1}{2f^{2}}((df)^{2}+(db+2BdA-2AdB)^{2}-
    \frac{1}{f}((dA)^{2}+(dB)^{2}))
    \label{eq:lineel}
\end{equation}
describes the invariant metric of the Riemannian symmetric space $S=
SU(2,1)/S[U(1,1)\times U(1)]$ in some coordinates. 
The stationary Einstein-Maxwell equations can thus be interpreted as 
three-dimensional gravity coupled to some matter model. The `matter' is 
a $SU(2,1)/S[U(1,1)\times U(1)]$ nonlinear sigma model 
\cite{forger,maisonsigma}. Note that sigma models 
are related to harmonic maps \cite{eels}. 
The space $S$ can be parametrized by trigonal $3\times3$ matrices $V$,
\begin{equation}
    V=\left(
    \begin{array}{ccc}
	\sqrt{f} & 0 & 0  \\
	i\sqrt{2}\Phi & 1 & 0  \\
	(b+i|\Phi|^{2})/\sqrt{f} & 
	(\sqrt{2}\bar{\Phi})/\sqrt{f} & 1/\sqrt{f}
    \end{array}\right)
    \label{group1}.
\end{equation}
The matrix $V$ satisfies 
\begin{equation}
    V^{\dagger}\eta V=\eta , \quad \eta =\left(
    \begin{array}{ccc}
	0 & 0 & i  \\
	0 & 1 & 0  \\
	-i & 0 & 0
    \end{array}\right)
    \label{eta},
\end{equation}
i.e.\ it is unitary with respect to the metric $\eta$ of $SU(2,1)$. 
The action of $\mathcal{G}\in SU(2,1)$ on $V$ is 
\begin{equation}
    V\to 
H(V,\mathcal{G})V\mathcal{G}^{-1}, \quad 
H(V,\mathcal{G})\in S[U(1,1)\times U(1)]
    \label{eq:suaction},
\end{equation}
where $H$ restores the triangular gauge of $V$. 
To obtain a gauge invariant parametrization, one introduces 
\begin{equation}
    \chi:= 
\Xi V^{\dagger}\Xi V, \quad \Xi=\mbox{diag}(1,-1,1)
    \label{eq:chi},
\end{equation}
on which the action of $\mathcal{G}\in SU(2,1)$ is given by 
\begin{equation}
    \chi 
\to \Xi(\mathcal{G}^{-1})^{\dagger}\Xi\chi \mathcal{G}^{-1}
    \label{eq:chiaction}.
\end{equation}
We have
\begin{equation}
    \chi=\left(
    \begin{array}{ccc}
	f-2|\Phi|^{2}+(b^{2}+|\Phi|^{4})/f & 
	\sqrt{2}\bar{\Phi} 
	(b-i|\Phi|^{2}+if)/f & (b-i|\Phi|^{2})/f  \\
	-\sqrt{2}\Phi(b+i|\Phi|^{2}-if)/f & 
	1-2|\Phi|^{2}/f & -\sqrt{2}\Phi/f  \\
	(b+i|\Phi|^{2})/f & \sqrt{2}\bar{\Phi}/f & 
	1/f
    \end{array}\right)
    \label{group2}.
\end{equation}
The $SU(2,1)$ symmetry can be used to generate solutions by the action of an 
element $\mathcal{G}$. We list the infinitesimal transformations  
and their consequences:\\

$\left(
\begin{array}{ccc}
    0 & 0 & 0  \\
    \theta_{1} & 0 & 0  \\
    \theta_{2} & \theta_{3} & 0
\end{array}\right)$ 

lead to gauge transformations which add physically 
irrelevant constants to $\Im \mathcal{E}$ and $\Im \Phi$, \\

$\left(
\begin{array}{ccc}
    0 & 0 & \theta  \\
    0 & 0 & 0  \\
    0 & 0 & 0
\end{array}\right)$ 

is an Ehlers transformation \cite{ehlers}
which changes $f\to b$, i.e.\ which generates stationary from static 
solutions (if the ADM-mass of the spacetime is non-zero, the 
transformed solution will have a Newman-Unti-Tambourini (NUT) 
parameter which corresponds to a magnetic monopole and which 
is believed to be unphysical) 

$\left(
\begin{array}{ccc}
    i\theta & 0 & 0  \\
    0 & -2i\theta & 0  \\
    0 & 0 & i\theta
\end{array}\right)$ 

is an electromagnetic duality transformation, i.e.\ the $U(1)$ 
symmetry of the sourcefree Maxwell equations,

$\left(
\begin{array}{ccc}
    \theta & 0 & 0  \\
    0 & 0 & 0  \\
    0 & 0 & -\theta
\end{array}\right)$ 

is a scale transformation, $f,b,\Phi\to 
e^{\theta}f, e^{\theta}b, e^{\theta/2}\Phi$, and 

$\left(
\begin{array}{ccc}
    0 & -i\theta & 0  \\
    0 & 0 & \bar{\theta}  \\
    0 & 0 & 0
\end{array}\right)$ 

is a Harrison transformation \cite{harrison} 
which changes $f\to \Phi$, i.e.\ generates solutions with electromagnetic 
fields from pure vacuum solutions.

\subsection{Harrison transformations}
Harrison transformations offer the possibility to generate solutions 
with charge from pure vacuum solution. 
This leads to  solutions to the 
Einstein-Maxwell equations containing one additional constant 
parameter which is related to the charge. For physical reasons we are 
interested in solutions which are equatorially symmetric 
and asymptotically flat, i.e.\
$f\to 1$, $\Phi\to0$ and $b\to0$ for $|\xi|\to 
\infty$. Asymptotic flatness is just the mathematical formulation of  
the physical concept of an isolated matter distribution, e.g.\ a 
galaxy. Solutions with equatorial symmetry, i.e.\ a class of solutions 
where the metric functions have a reflection symmetry at the 
equatorial plane $\zeta=0$, are of special physical interest. 
In a Newtonian setting it can be proven that perfect fluids 
in thermodynamical equilibrium 
lead to equatorially symmetric situations, and the same is assumed to 
hold in a general relativistic context. A consequence of this 
condition is that NUT-parameters are ruled out.

We assume that the pure vacuum solutions which we want to 
submit to a Harrison transformation satisfy these conditions.
To ensure that the transformed solutions have the same 
asymptotic behavior, one has to use a scale transformation ($f\to1$)
together with 
a transformation which changes $\Phi$ and $b$ by some constant 
($\Phi,b\to 0$). By 
exponentiating the matrices of the $SU(2,1)$ transformations, we 
thus consider a transformation of the form 
\begin{equation}
    \mathcal{G}=\left(
    \begin{array}{ccc}
	1 & i\theta_{1} & -i\theta_{1}\bar{\theta_{1}}/2  \\
        0 & 1 & -\bar{\theta_{1}}  \\
        0 & 0 & 1
    \end{array}\right)\mbox{diag}(e^{-\theta_{2}},1,e^{\theta_{2}})
    \left(
    \begin{array}{ccc}
        1 & 0 & 0  \\
        -\theta_{3} & 1 & 0  \\
        -\theta_{4}+\theta_{3}\theta_5/2 & -\theta_5 & 1
    \end{array}\right).
    \label{group6}
\end{equation}
Since the asymptotic 
conditions imply that $\chi$ and $\chi'$ are  the 
unit matrix at infinity, the matrix $\mathcal{G}$ must satisfy the condition 
 $\Xi \mathcal{G}^{\dagger}\Xi =\mathcal{G}^{-1}$. 
This leads with (\ref{group6}) to
\begin{equation}
    e^{\theta_2}=\frac{1}{1-\theta_1\bar{\theta_1}/2},\quad 
    \theta_3=i\bar{\theta_1}, \quad \theta_4=0,\quad \theta_5=\theta_1
    \label{group9}.
\end{equation}
The matrix $\mathcal{G}$ thus takes the form
\begin{equation}
    \mathcal{G}=\frac{1}{1-\theta_1\bar{\theta_1}/2}\left(
    \begin{array}{ccc}
        1 & i\theta_1 & -i\theta_1\bar{\theta_1}/2  \\
        -i\bar{\theta_1} & 1+\theta_1\bar{\theta_1}/2 & -\bar{\theta_1}  \\
         i\theta_1\bar{\theta_1}/2& -\theta_1 & 1
    \end{array}
    \right)
    \label{group10}.
\end{equation}
If we transform an Ernst potential of a pure vacuum solution 
($\Phi=0$), we end up with 
\begin{equation}
    \Phi'=-\frac{\theta_1}{\sqrt{2}}\frac{\theta_1\bar{\theta_1}(f^{2}+b^{2})
    /2-(1+\theta_1\bar{\theta_1}/2)f+1-ib(1-\theta_1\bar{\theta_1}/2)}{(1-\theta_1\bar{\theta_1}f/2 
    )^{2}+(\theta_1\bar{\theta_1}/2)^{2}b^{2}}
    \label{group11}.
\end{equation}
We are interested in transformations which preserve the equatorial
symmetry, i.e.\ $f(-\zeta)=f(\zeta)$, $b(-\zeta)=-b(\zeta)$ and 
$\Phi(-\zeta)=\bar{\Phi}(\zeta)$. This implies for (\ref{group11}) 
that $\theta_1$ must be real which rules out magnetic monopoles. We 
put $q=\theta_1/\sqrt{2}$ and sum up the results for the transformed 
potentials:
\begin{eqnarray}
    f' & = & \frac{(1-q^{2})^{2}f}{(1-q^{2}f)^{2}+q^{4}b^{2}}
    \label{f'} , \\
    b' & = & \frac{(1-q^{4})b}{(1-q^{2}f)^{2}+q^{4}b^{2}}
    \label{b'} , \\
    \Phi' & = & -q\frac{(1-f)(1-q^{2}f)+q^{2}b^{2} +ib(1-q^{2})}{
    (1-q^{2}f)^{2}+q^{4}b^{2}}
    \label{Phi'}.
\end{eqnarray}
The real parameter $q$ has to be in the region $0<|q|<1$, for $q>1$ 
the transformed spacetime would have a negative mass if the original 
mass was positive. The value 
$q=0$ corresponds to the untransformed solution. The above formulas 
imply that the functions $f'$, $b'$ and $\Phi'$ are analytic 
where the original functions are analytic.

A well-known example is the Harrison
transformation of the Schwarzschild 
solution which leads to the Reissner-Nordstr\"om solution. In the 
Ernst picture the Schwarzschild solution reads 
in cylindrical Weyl coordinates 
with  $r_{\pm}=\sqrt{(\zeta\pm m)^{2}+\rho^{2}}$
\begin{equation}
    f=\frac{r_{+}+r_{-}-2m}{r_{+}+r_{-}+2m}, \quad b=0
    \label{eq:scharschild},
\end{equation}
where the horizon ($f=0$) is located on the axis between $-m$ and 
$m$. For the transformed solution we get with (\ref{Phi'}) 
\begin{equation}
    f'  =  \frac{(r_{+}+r_{-})^{2}-4m^{2}}{(r_{+}+r_{-}+2m')^{2}},
   \quad
    \Phi'  =  \frac{2Q}{r_{+}+r_{-}+2m'}, \quad m' =m\frac{1+q^{2}}{
    1-q^{2}}, \quad Q = -\frac{2mq}{1-q^{2}}
    \label{reissner},
\end{equation}
and $b'=0$ which is the Reissner-Nordstr\"om solution. 
This is a static 
spacetime with mass $m'$ and charge $Q$ subject to the 
relation $m'{}^{2}-Q^{2}=m^{2}$. 
Both $m'$ and $Q$ diverge for $q\to1$. The extreme Reissner-Nordstr\"om 
solution with $m'=Q$ is only possible in the limit $m\to0, |q|\to1$.
The horizon of the solution is again located on the axis between $\pm 
m$ which illustrates that the horizon degenerates in the extreme case. 

\subsection{Asymptotic behavior of the Harrison transformed solutions}
We assume that the asymptotic behavior of the original solution, 
which can be read off on the axis, is of the form 
$f=1-2M/|\zeta|$, $b=-2J/\zeta^{2}$ and $\Phi=Q/|\zeta|-iJ_{M}/\zeta^{2}$ plus 
terms of lower order in $1/|\zeta|$ where 
$M$ is the Arnowitt-Deser-Misner mass, $J$ the angular momentum, $Q$ 
the electric charge and $J_{M}$ the magnetic moment.  The same will 
hold for the Harrison transformed potentials. 
We find \cite{prharrison}
\begin{equation}
    M'=M\frac{1+q^{2}}{1-q^{2}}-\frac{2q}{1-q^{2}}Q,\quad 
    J'=J\frac{1+q^{2}}{1-q^{2}}-\frac{2q}{1-q^{2}}J_{M},
    \label{group12}
\end{equation}
and
\begin{equation}
    Q'=Q\frac{1+q^{2}}{1-q^{2}}-\frac{2q}{1-q^{2}}M,\quad 
    J_{M}'=J_{M}\frac{1+q^{2}}{1-q^{2}}-\frac{2q}{1-q^{2}}J
    \label{group12a}.
\end{equation}
It is interesting to note that the quantities $M^{2}-Q^{2}$ and $J^{2}
-J_{M}^{2}$ are invariants of the transformation. They are related to 
the Casimir operator of the $SU(2,1)$-group.
If the original solution was uncharged, 
the extreme relation $M'=\pm Q'$ is only possible 
in the limit $M\to 0$. 

A consequence of the relations (\ref{group12a}) is the presence of a 
non-vanishing charge if the ADM mass of the original solution is 
non-zero whereas the charge is. Since charges normally compensate in 
astrophysical settings, this limits the astrophysical relevance of 
Harrison transformed solutions. 

A further invariant is the combination $J_{M}M-JQ$ which is of 
importance in relation to the gyromagnetic ratio 
\begin{equation}
    g_{M}=\frac{2MJ_{M}}{JQ}.
    \label{gyro}
\end{equation}
Relation (\ref{group12}) implies that $g_{M}'$ is 
equal to 2 if $Q=J_{M}=0$ and $q\neq 0$. Thus all 
solutions which can be generated via a Harrison transformation from 
solutions with vanishing electromagnetic fields as the Kerr-Newman 
family from Kerr
have a gyromagnetic ratio of 2. Due to the invariance of 
$J_{M}M-JQ$ under Harrison transformations, a gyromagnetic 
ratio of 2 is not changed under the transformation.  

Whether this 
property is an indication of a deep relation between relativistic 
quantum mechanics and general relativity as claimed in 
\cite{kipfi02,kipfi03} is an open question. Here it is just related 
to an invariant of the Harrison transformation, a subgroup of 
$SU(2,1)$. Since most of the 
known solutions to the Einstein-Maxwell equations can be 
generated via a 
Harrison transformation from solutions to the pure vacuum equations as 
the Kerr-Newman family from Kerr, a gyromagnetic ratio of 2 is well 
known from exact solutions. Numerical calculations of charged neutron 
stars \cite{jerome} indicate, however, that values well below 2 are 
to be expected in astrophysically realistic situations. 

\subsection{The stationary axisymmetric case}
In the astrophysically important stationary axisymmetric case, the 
symmetry group of the equations increases again, this time to the 
infinite dimensional Geroch group \cite{gerochgroup,breimai}. This means that 
the equations are completely integrable, where the notion of 
integrability is to be understood in a Hamiltonian sense: the 
equations have the same number of conserved quantities as degrees of 
freedom. For completely integrable systems, this number tends in a 
countable way to 
infinity. The infinite dimensional symmetry group shows up in
treating the differential equation under consideration as the 
integrability condition of an overdetermined linear differential 
system for a matrix-valued function. 
The system 
contains an additional parameter, the so-called spectral parameter 
which reflects the infinite dimensional symmetry group. 

In the presence of a second Killing vector, the metric 
(\ref{maison1}) can be further specialized. The axial Killing vector 
$\partial_{\phi}$ commutes with the timelike Killing vector 
$\partial_{t}$. The metric $h_{ab}$ of (\ref{maison1})
can be chosen to be diagonal,
$h_{ab}=\mbox{diag}(e^{2k},e^{2k},\rho^{2})$, and $k_{a}$ can be brought into 
the form $k_{a}=(0,0,a)$ which leads to the Weyl-Lewis-Papapetrou 
metric, see e.g.\ \cite{exac}
\begin{equation}
\mathrm{ d} s^2 =-e^{2U}(\mathrm{ d} t+a\mathrm{ d} \phi)^2+e^{-2U}
\left(e^{2k}(\mathrm{ d} \rho^2+\mathrm{ d} \zeta^2)+ \rho^2\mathrm{
d} \phi^2\right), \label{vac1}
\end{equation}
where $f=e^{2U}$. 
In this case the Einstein-Maxwell
equations reduce to 
\begin{eqnarray}
    f\Delta \mathcal{E} & = & (\nabla \mathcal{E}+2\bar{\Phi} \nabla 
    \Phi)\nabla \mathcal{E},
    \nonumber  \\
    f\Delta \Phi & = & (\nabla \mathcal{E}+2\bar{\Phi} \nabla 
    \Phi)\nabla \Phi,
    \label{ernst}
\end{eqnarray}
where $\Delta$ and $\nabla$ are the standard differential operators in 
cylindrical coordinates, and where the potentials $\mathcal{E}$ and $\Phi$ 
are independent of $\phi$. The first equation generalizes the Newtonian 
equation $\Delta U=0$ to the Einstein-Maxwell case, the 
second equation generalizes the 
Maxwell equation (\ref{eq:max2}). 
The duality relations (\ref{eq:B2}) and (\ref{eq:twist}) read 
\begin{eqnarray}
    (\Im \Phi)_{\xi}&=&\frac{i}{\rho}f
    (A_{\phi,\xi}-aA_{t,\xi})
	\label{magnetic},
     \\
    a_{\xi} & = & \frac{\rho}{f^{2}}\left(\mathrm{i}(\Im 
    \mathcal{E})_{\xi} +\Phi\bar{\Phi}_{\xi}-\bar{\Phi}\Phi_{\xi}\right)
    \label{a},
\end{eqnarray}
which implies that $a$ and $A_{3}$ follow 
from $\mathcal{E}$ and $\Phi$. We choose a gauge where $A_{1}=A_{2}=0$.
The equations for $\mathcal{R}_{ab}$ of (\ref{eq:actint})
are equivalent to
\begin{equation}
 k_{\xi}  =  \frac{\xi-\bar{\xi}}{f}\left( \frac{1}{4f}(
    \mathcal{E}_{\xi} 
    +2\bar{\Phi}\Phi_{\xi} )(\bar{\mathcal{E}}_{\xi} 
    +2\Phi\bar{\Phi}_{\xi} ) - \Phi_{\xi}\bar{\Phi}_{\xi}\right)
    \label{k}.
\end{equation}
Thus the complete metric and the electromagnetic potential can be 
obtained from given potentials $\mathcal{E}$ and $\Phi$ via quadratures.

The system (\ref{ernst}) was shown to be completely 
integrable in  \cite{meudon}. 
In the form \cite{neukramer}, the associated 
linear differential system for a $3\times 3$ 
matrix-valued function $\Psi$ reads (we use the complex coordinate
$\xi=\zeta-i\rho$)
\begin{eqnarray}
    \Psi_{\xi}\Psi^{-1} & = & \left( 
    \begin{array}{ccc}
	\mathcal{D}_1 & 0 & \mathcal{M}_1  \\
	0 & \mathcal{C}_1 & 0  \\
	-\mathcal{N}_1 & 0 & \frac{1}{2}(\mathcal{C}_1+\mathcal{D}_1)
    \end{array}
    \right)
    +\frac{K-\bar{\xi}}{\mu_{0}}\left( 
    \begin{array}{ccc}
	0 & \mathcal{D}_1 &  0 \\
	\mathcal{C}_1 & 0 & -\mathcal{M}_1  \\
	0 & -\mathcal{N}_1 & 0
    \end{array}
    \right)
    \nonumber , \\
    \Psi_{\bar{\xi}}\Psi^{-1} & = & \left( 
    \begin{array}{ccc}
	\mathcal{D}_2 & 0 & \mathcal{M}_2  \\
	0 & \mathcal{C}_2 & 0  \\
	-\mathcal{N}_2 & 0 & \frac{1}{2}(\mathcal{C}_2+\mathcal{D}_2)
    \end{array}
    \right)
    +\frac{K-\xi}{\mu_{0}}\left( 
    \begin{array}{ccc}
	0 & \mathcal{D}_2 &  0 \\
	\mathcal{C}_2 & 0 & -\mathcal{M}_2  \\
	0 & -\mathcal{N}_2 & 0
    \end{array}
    \right)
    \label{linear},
\end{eqnarray}
where $\Psi$ depends on the spectral parameter $K$ which varies on 
the Riemann surface $\mathcal{L}$ 
of genus zero given by the relation 
$\mu_{0}^{2}(K)=(K-\bar{\xi})(K-\xi)$. This is a first 
hint on the relevance of Riemann surfaces in this context. 
Notice the special feature of the 
Ernst equation 
that the branch points $\xi$, $\bar{\xi}$ depend on the spacetime 
coordinates. 
We denote a 
point on $\mathcal{L}$ with the projection $K$ in the complex plane by 
$P=(K,\pm \mu_{0}(K))=K^{\pm}$. On $\mathcal{L}$ there is an
involution $\sigma$ that interchanges the sheets, i.e.\ with 
$P=(K,\mu_{0}(K))$ we have $\sigma P= P^{\sigma}=(K,-\mu_{0}(K))$.
We use the
notation $\infty^{\pm}$ for the infinite points on different sheets of
the curve $\mathcal{L}$, namely $\mu/K^{g+1}\to \pm 1$ as $K\to
\infty^{\pm}$.

The expressions 
for $\mathcal{C}_{i}$, $\mathcal{D}_{i}$ and $\mathcal{M}_{i}$ 
($i=1,2$) follow from the condition
\begin{equation}
    \Psi(\infty^{+},z,\bar{z})=\left( 
    \begin{array}{ccc}
	\bar{\mathcal{E}}+2\Phi\bar{\Phi} & 1 &  \sqrt{2}\mathrm{i}\Phi \\
	\mathcal{E} & -1 & -\sqrt{2}\mathrm{i}\Phi  \\
	-2\mathrm{i}\bar{\Phi}e^{U} & 0 & \sqrt{2}e^{U}
    \end{array}
    \right)=\left( 
    \begin{array}{ccc}
	1 & 0 &  \sqrt{2}\mathrm{i}\Phi \\
	0 & 1 & 0  \\
	0 & 0 & \sqrt{2}e^{U}
    \end{array}
    \right)\left( 
    \begin{array}{ccc}
	\bar{\mathcal{E}} & 1 &  0 \\
	\mathcal{E} & -1 & -\sqrt{2}\mathrm{i}\Phi  \\
	-\sqrt{2}\mathrm{i}\bar{\Phi} & 0 & 1
    \end{array}
    \right),
    \label{infinity}.
\end{equation}
Thus we have $\det 
\Psi(\infty^{+})=-2\sqrt{2}e^{3U}$. The reality conditions read
\begin{equation}
    \mathcal{C}_2=\bar{\mathcal{D}}_{1}, \quad 
    \mathcal{C}_1=\bar{\mathcal{D}}_{2}, \quad 
    \mathcal{N}_2=-\bar{\mathcal{M}}_{1}, \quad \mathcal{M}_2=-\bar{\mathcal{N}}_{2}.
    \label{real}
\end{equation}
The inverse matrix takes the form
\begin{equation}
    \Psi^{-1}=\frac{1}{2f}\left(
    \begin{array}{ccc}
	1 & 1 & 0  \\
	\mathcal{E}+2\Phi\bar{\Phi} & -\bar{\mathcal{E}} & -2i\Phi e^{U}  \\
	\sqrt{2}i\bar{\Phi} & \sqrt{2}i\bar{\Phi} & \sqrt{2}e^{U}
    \end{array}
    \right)
    \label{psi-1},
\end{equation}
which implies via (\ref{linear})
\begin{eqnarray}
    \mathcal{C}_1 & = & (\mathcal{E}_{\xi}+2\bar{\Phi}\Phi_{\xi})/(2f)
    \nonumber,  \\
    \mathcal{D}_1 & = & (\bar{\mathcal{E}}_{\xi}+2\Phi\bar{\Phi}_{\xi})/(2f)
    \nonumber , \\
    \mathcal{M}_1 & = & i\Phi_{\xi}e^{-U}
    \nonumber , \\
    \mathcal{N}_1 & = & i\bar{\Phi}_{\xi}e^{-U}
    \label{linear2}.
\end{eqnarray}

The system (\ref{linear}) can be used to generate solutions to the 
Ernst equations. To this end one investigates the singularity 
structure of the matrices $\Psi_{\xi}\Psi^{-1}$ and $\Psi_{\bar{\xi}} 
\Psi^{-1}$ with respect to the spectral parameter and infers a set of 
conditions for the matrix $\Psi$ which satisfies the linear system 
(\ref{linear}). These conditions can be summarized in the following 
theorem (see \cite{koro88}):\\
\begin{theorem}
    The matrix $\Psi$ is at least twice differentiable with respect to 
    $\xi$ and $\bar{\xi}$. $\Psi(P)$ is holomorphic and invertible at the branch 
    points $\xi$ and $\bar{\xi}$ such that the logarithmic derivative
    $\Psi_{\xi}\Psi^{-1}$ has a pole at $\xi$ and 
    $\Psi_{\bar{\xi}}\Psi^{-1}$ has a pole  at $\bar{\xi}$.\\
    II. All singularities of $\Psi$ on $\mathcal{L}$  
    are such that the logarithmic derivatives 
    $\Psi_{\xi}\Psi^{-1}$ and $\Psi_{\bar{\xi}}\Psi^{-1}$ are 
    holomorphic there. \\
III.  The matrix function $\Psi$ is 
subject to the reduction condition 
\begin{equation}
    \epsilon \Psi(P^{\sigma})=\Psi(P)\gamma(P)
    \label{reduction},
\end{equation}
where $\epsilon=\mbox{diag}(1,-1,1)$, and where $\gamma$ is an 
invertible matrix 
independent of $\xi$, $\bar{\xi}$.\\
IV. At $\infty^{+}$, the matrix function $\Psi$ is given by 
(\ref{infinity}).   
\end{theorem}
A proof of this theorem may be obtained by comparing the matrix 
$\Psi$ with the linear system (\ref{linear}). 
\begin{proof}
    Because of I, $\Psi$ and $\Psi^{-1}$ can be expanded in a series in 
    the local parameters 
    $\tau_{\xi}=\sqrt{K-\xi}$ and $\tau_{\bar{\xi}}
    =\sqrt{K-\bar{\xi}}$ in a neighborhood of $P=\xi$
    and $P=\bar{\xi}\neq \xi$ respectively at all points $\xi$, $\bar{\xi}$ which 
    are not singularities of $\Psi$. This implies that $\Psi_{\xi}
    \Psi^{-1}=\alpha_0/t+\alpha_1+\alpha_2t +\ldots$. We recognize that, because of
    I and II, $\Psi_{\xi} \Psi^{-1}-\alpha_0/t$ is a holomorphic function. The
    normalization condition IV implies that this quantity is bounded at 
    infinity. According to Liouville's theorem, it is a constant. 
    Since $\Psi$, $\Psi^{-1}$ and $\Psi_{\xi}$ are single valued functions on $\mathcal{L}$, 
    they must be functions of $K$ and $\mu_{0}$. Therefore we have $\Psi_{\xi}
    \Psi^{-1}=\beta_0 \sqrt{\frac{K-\bar{\xi}}{K-\xi}}+\beta_1$. The matrix 
    $\beta_0$ must be 
    independent of $K$ and $\mu_{0}$ since $\Psi_{\xi} \Psi^{-1}$ must have the same 
    number of zeros and poles on $\mathcal{L}$. The structure of the matrices 
    $\beta_0$ and $\beta_1$ follows from III. From the normalization condition IV,
    it follows that $\Psi_{\xi} \Psi^{-1}$ has the structure of 
    (\ref{linear}).
    The corresponding equation for $\Psi_{\bar{\xi}}\Psi^{-1}$ can be obtained 
    in the same way.
\end{proof}

The choice of the matrix $\gamma$ in (\ref{lin7}) corresponds to 
a gauge freedom which is however not completely fixed. 
A matrix $C(K)$ with the property $[C,\gamma]=0$
with $C(\infty)=\hat{1}$ which acts on $\Psi$ in the form $\Psi\to \Psi 
C(K)$ leads again to a solution of (\ref{linear}) for the same 
potentials $\mathcal{E}$ and $\Phi$.

We note that the metric function $a$ can be directly obtained from the matrix $\Psi$ 
without integrating (\ref{a}): 
We denote by $D_{P} F(P)$ the 
coefficient of the linear term in the expansion of the function 
$F$ in the local parameter near $P$. Using the 
identity 
$(\Psi^{-1}D_{\infty^{+}}\Psi)_{\xi}=\Psi^{-1}D_{\infty^{+}}
(\Psi_{\xi}\Psi^{-1})\Psi$, 
one finds with (\ref{linear}) to (\ref{linear2})
\begin{equation}
    ((\Psi^{-1}D_{\infty^{+}}\Psi)_{\xi})_{12}=
    -\frac{i\rho}{2f}(\mathcal{C}_1-\mathcal{D}_1) =
    -\frac{i\rho}{2f^{2}}(i(\Im 
    \mathcal{E})_{\xi}+\bar{\Phi}\Phi_{\xi}-\Phi\bar{\Phi}_{\xi})
    \label{linear3}.
\end{equation}
Equation (\ref{a}) leads to the expression
\begin{equation}
    (a-a_{0})e^{2U}=iD_{\infty^{+}}(\Psi_{12}+\Psi_{22})
    \label{A2},
\end{equation}
which relates the metric function $a$ directly to the potential $\Psi$; 
the constant $a_{0}$ is determined by the condition that $a$ has to 
vanish on the regular part of the axis.

To generate solutions to the Ernst equation one has to construct 
matrices according to the above theorem. The theorem ensures that the 
so constructed 
solutions to the Ernst equation are analytic at all points where the 
conditions I-IV are satisfied. It also makes clear where 
singularities of the spacetime can be found: if the moving branch 
points $\xi$, $\bar{\xi}$ coincide with singularities of $\Psi$, 
condition I is no longer fulfilled and the theorem does not hold. 
These points are possible singularities, but they could 
be regular if the singularities of $\Psi$ are for instance pure 
gauge. On the other hand, if the singularities of the spacetime are 
prescribed --- for electro-vacuum the boundary of some matter 
source constitutes 
a singularity --- one has to construct a matrix with the 
corresponding singularity. This means one has to solve a so-called 
Riemann-Hilbert problem, i.e.\ to find a function with prescribed 
singularities on $\mathcal{L}$. 

This method to construct solutions to integrable 
equations is also known as the inverse scattering method \cite{soliton}.
Explicit solutions are in general only known for scalar 
Riemann-Hilbert problems which lead in the case of the Ernst equation 
to static solutions for the pure vacuum, see \cite{koro88,prd,jgp2}. In this 
case the Ernst equation reduces to the axisymmetric Laplace equation, 
the Euler-Darboux equation. In the matrix case, the solution to a 
Riemann-Hilbert problem is equivalent to an integral equation, see 
\cite{alekseev,prd,exac} for the Ernst equation. A simple special case are 
so-called `soliton'-solutions or B\"acklund transformations where the 
matrix $\Psi$ has only poles and zeros of the determinant as singularities. 
Since the Ernst equation is an elliptic equation, it has obviously no 
solutions describing physical solitons, but it has solutions with the 
same mathematical properties. The most prominent representant of this 
class for the Ernst equation is the Kerr-Newman family of charged 
rotating black holes. 

In \cite{prd} it was shown for the pure vacuum case
that the Riemann-Hilbert problem can be 
solved explicitly for a large class of problems by exploiting the 
gauge freedom of the matrix $\Psi$. This leads to solutions which are 
defined on certain Riemann surfaces. 
It is well known that large classes of solutions to non-linear integrable equations 
can be constructed via methods from algebraic geometry. For evolution 
equations like Korteweg-de Vries and Sine-Gordon, see e.g.\ \cite{algebro}, 
the corresponding solutions are periodic or quasi-periodic. 
Algebro-geometric solutions to the Ernst equation do not show any 
periodicity as will be discussed in the following section. A way to 
construct such solutions is via the so-called monodromy matrix 
\cite{itogi,algebro}. For a 
linear system of the form 
\begin{equation}
    \Psi_{\xi}=W\Psi, \quad \Psi_{\bar{\xi}}=V\Psi
    \label{eq:lin2}
\end{equation}
as (\ref{linear}), the monodromy matrix $L$ can be defined as the 
solution of the linear differential system 
\begin{equation}
    L_{\xi}=[W,L], \quad L_{\bar{\xi}}=[V,L]
    \label{eq:monodromy}.
\end{equation}
It follows from equation (\ref{eq:monodromy}) that the characteristic 
polynomial 
\begin{equation}
    Q(\hat{\mu},K)=\det(L-\hat{\mu}\hat{1})
    \label{eq:char}
\end{equation}
is independent of the physical coordinates (the coefficients of the 
polynomial are `integrals of motion'). The equation  $Q(\hat{\mu},K)=0$ 
is then the equation of a plane algebraic curve. Since $L$ is a 
$3\times3$-matrix, relation (\ref{eq:char}) is cubic in $\hat{\mu}$ 
which can be always brought into normal form by a redefinition of 
$\hat{\mu}$: 
\begin{equation}
    \hat{\mu}^{3}+\mathcal{P}(K)\hat{\mu}+\mathcal{Q}(K)=0
    \label{eq:mueq}.
\end{equation}
The functions $\mathcal{P}$ and $\mathcal{Q}$ are analytic in $K$. 
Equation (\ref{eq:mueq}) defines a three-sheeted Riemann surface which 
will  in general have infinite genus. For polynomial $\mathcal{P}$ 
and $\mathcal{Q}$, the surface will be compact and will have finite 
genus. For a given surface the solutions to the Ernst equations can be 
given in terms of the theta functions on this surface which was first 
done in \cite{koro88} for a special case. 

Since the theory of these 
surfaces is not as well understood as the theory of hyperelliptic surfaces 
which occur in the pure vacuum case, we will study in the next section 
Harrison transformed hyperelliptic solutions. In section 6, we will 
come back to solutions 
to the Ernst equations on three-sheeted surfaces. 

\subsection{Conformastationary and magnetostatic solutions}
There are two special cases of the above equations
for which disk sources 
have been considered: Conformastationary metrics \cite{perjes,iswi} 
are included in the 
above formalism as metrics with a flat metric $h_{ab}$, there is no 
axial symmetry required. Writing 
$f=(V\bar{V})^{-1}$, the function $V$ is a complex solution to the 
three-dimensional Laplace equation. For asymptotically flat solutions 
$V$ has to tend to 1 at infinity. The electromagnetic fields are given 
by 
\begin{equation}
    \mathbf{E}+i\mathbf{H}=\mbox{grad}\left(\frac{1}{V}\right),\quad
    \mathbf{D}+i\mathbf{B}=|V|(\mathbf{E}+i\mathbf{H})+i\mathbf{T}
    \times (\mathbf{E}+i\mathbf{H}),
    \label{eq:conforma}
\end{equation}
where $\mathbf{T}$ is a solution to 
\begin{equation}
    \mbox{rot}\mathbf{T}=i(V\mbox{grad}\bar{V}-\bar{V}\mbox{grad}V)
    \label{eq:conforma2}.
\end{equation}

Static spacetimes with a pure magnetic field can be mapped to the case 
of stationary axisymmetric vacuum spacetimes, see \cite{exac,letelier}. Writing 
$A_{\mu}=(0,0,0,A(\rho,\zeta))$, the field equations reduce to the 
Ernst equation which is discussed in the next section. Thus all 
solutions to the vacuum Ernst equation can be interpreted as magnetostatic 
solutions by putting $f\to 1/(\Re \mathcal{E})^{2}$ and $A \to
-\sqrt{2}\Im \mathcal{E}$.

\section{Stationary axisymmetric Einstein equations, theta functional 
solutions and counter-rotating dust disks 
under Harrison transformations}
In this section we will  summarize results on the 
Ernst equation in the pure vacuum case, 
a class of solutions on hyperelliptic 
Riemann surfaces \cite{koro88} and a member of this class 
describing counter-rotating dust disks \cite{prl2} 
which were discussed in \cite{prd3} and \cite{prd4}. We study the 
action of a Harrison transformation on stationary axisymmetric 
solutions and discuss the transformed counter-rotating dust disk as 
an example. 
\subsection{The stationary axisymmetric vacuum}
In the stationary axisymmetric vacuum, the Einstein-Maxwell equations 
of the previous section hold with $\Phi=0$. Since the simplification 
is considerable, we list the relevant equations below. The Ernst 
potential $\mathcal{E}=f+ib$ is subject to the Ernst equation 
\cite{ernst1}
\begin{equation}
\mathcal{E}_{\xi\bar{\xi}}-\frac{1}{2(\bar{\xi}-\xi)}(
\mathcal{E}_{\bar{\xi}}-
\mathcal{E}_{\xi})=\frac{2
}{\mathcal{E}+\bar{\mathcal{E}}} \mathcal{E}_{\xi} 
\mathcal{E}_{\bar{\xi}} \label{vac10}.
\end{equation}
If the 
Ernst potential is real, equation (\ref{vac10}) is equivalent to 
$\Delta U=0$. The corresponding spacetime is static and 
belongs to the so-called Weyl class. For a 
given Ernst potential the metric
(\ref{vac1}) follows from 
\begin{eqnarray}
     a_{\xi}&=&2\rho\frac{(\mathcal{E}-\bar{\mathcal{E}})_{\xi}}{
     (\mathcal{E}+\bar{\mathcal{E}})^{2}}\label{vac9},\\
k_{\xi}&=&(\xi-\bar{\xi}) \frac{\mathcal{E}_{\xi}\bar{\mathcal{E}}_{\xi}}{(
\mathcal{E}+\bar{\mathcal{E}})^{2}} \label{2.10a10}.
\end{eqnarray}

The stationary axisymmetric Einstein equations in vacuum were shown 
to be completely integrable in \cite{maison} and \cite{belzak}. 
The associated linear system can be formulated for 
a $2\times 2$-matrix $\Psi$ (see \cite{korot92})
\begin{eqnarray}
    \Psi_{\xi}\Psi^{-1} & = & \left(
    \begin{array}{cc}
	\mathcal{M} & 0  \\
	0 & \mathcal{N}
    \end{array}
    \right)+\sqrt{\frac{K-\xi}{K-\bar{\xi}}}\left(
    \begin{array}{cc}
	0 & \mathcal{M}  \\
	\mathcal{N} & 0
    \end{array}\right)
    \nonumber, \\
\Psi_{\bar{\xi}}\Psi^{-1} & = & \left(
    \begin{array}{cc}
	\bar{\mathcal{N}} & 0  \\
	0 & \bar{\mathcal{M}}
    \end{array}
    \right)+\sqrt{\frac{K-\xi}{K-\bar{\xi}}}\left(
    \begin{array}{cc}
	0 & \bar{\mathcal{N}}  \\
	\bar{\mathcal{M}} & 0
    \end{array}\right)
    \label{a1},
\end{eqnarray}
where
\begin{equation}
    \mathcal{M}=\frac{\bar{\mathcal{E}}_{\xi}}{\mathcal{E}+\bar{\mathcal{E}}},
    \quad \mathcal{N}=\frac{\mathcal{E}_{\xi}}{\mathcal{E}+\bar{\mathcal{E}}}
    \label{MN}.
\end{equation}

Theorem 3.1 holds with the following changes:\\
\emph{III'. $\Psi$ is subject to the reduction condition
\begin{equation}
	\Psi(P^{\sigma}) = \sigma_3 \Psi(P) \sigma_2
	\label{lin7},
\end{equation}
where  $\sigma_2$, $\sigma_{3}$ are Pauli matrices.\\
IV'. The normalization and reality condition
\begin{equation}
	\Psi(P=\infty^1)=\left(
	\begin{array}{cc}
		\bar{\mathcal{E}} & -i  \\
		\mathcal{E} & i
	\end{array}
	\right)
	\label{lin9}.
\end{equation}
The function $\mathcal{E}$ in (\ref{lin9}) is then a solution to the 
Ernst equation (\ref{vac10}).}

We note that the choice of the normalization of $\Psi$ at infinity is 
different from the one in theorem 3.1. This form was chosen to 
implement the Harrison transformation as was done in the last section 
with the metric $\eta$ in (\ref{eta}) of $SU(2,1)$.
The choice of the matrix $\sigma_{2}=\left(
\begin{array}{cc}
    0 & -i  \\
    i & 0
\end{array}\right)$ in condition III is again a gauge condition, which 
does not fix the gauge completely, however. The remaining gauge 
freedom is here due to matrices $C(K)=\kappa_{1}(K) \hat{1}+\kappa_{2}(K) 
\sigma_{2}$, where the $\kappa_{i}$ do not depend on $\xi$, 
$\bar{\xi}$ and obey the asymptotic conditions $\kappa_{1}(\infty)=1$ and 
$\kappa_{2}(\infty)=0$. The matrices $C$ act on $\Psi$ 
in the form $\Psi\to \Psi C(K)$. 
Since it is a consequence of (\ref{a1}) that $\det \Psi=F(K) e^{2U}$ 
where $F(K)$ 
is independent of $\xi$, $\bar{\xi}$ 
we can use this gauge freedom to 
choose $F(K)=1$.
The linear system (\ref{a1}) leads for the matrix $\chi(P)=
\Psi^{-1}(\infty^{-})\Psi(P)$ to one of the linear systems 
used in \cite{breimai}. This parametrization, especially
\begin{equation}
    \Psi^{-1}(\infty^{-})\Psi(\infty^{+})=\frac{1}{\mathcal{E}+
    \bar{\mathcal{E}}}\left(
    \begin{array}{cc}
	2\mathcal{E}\bar{\mathcal{E}} & 
	i(\bar{\mathcal{E}}-\mathcal{E})  \\
	 i(\bar{\mathcal{E}}-\mathcal{E}) & 2
    \end{array}\right)
    \label{a3},
\end{equation}
reveals that the Ernst equation is an  $SL(2,\mathbb{R})/SO(2)$ sigma 
model. For more details on the group aspect see 
\cite{breimai} and the 
discussion in the previous section.

For given  $\Psi(K)$, one can again
directly determine the metric function $a$ without having to 
integrate relation (\ref{vac9}).  With (\ref{a1}) one finds for 
the matrix $\mathcal{S}=
\Psi^{-1}(\infty^{+})D_{\infty^{+}}\Psi(\infty^{+})$
\begin{equation}
   \mathcal{S}_{\xi}:= (\Psi^{-1}(\infty^{+})D_{\infty^{+}}\Psi)_{\xi}=\frac{\xi-
    \bar{\xi}}{2(\mathcal{E}+\bar{\mathcal{E}})^{2}}\left(
    \begin{array}{cc}
	(\mathcal{E}\bar{\mathcal{E}})_{\xi} & i(\bar{\mathcal{E}}-
	\mathcal{E})_{\xi}\\
	i(\mathcal{E}^{2}\bar{\mathcal{E}}_{\xi}-\bar{\mathcal{E}}^{2}
	\mathcal{E}_{\xi})& -(\mathcal{E}\bar{\mathcal{E}})_{\xi}
    \end{array}\right)
    \label{a4}.
\end{equation}
This implies with (\ref{vac9})
\begin{equation}
    a-a_{0}=-2\mathcal{S}_{12}=-2(\Psi^{-1}(\infty^{+})D_{\infty^{+}}\Psi)_{12},
    \label{a5}
\end{equation}
where $a_{0}$ is a constant which is fixed by the condition 
that $a$ vanishes on the regular part of the axis. The matrix $\mathcal{S}$ 
will be needed to calculate the metric function $a$ after a 
Harrison transformation. 

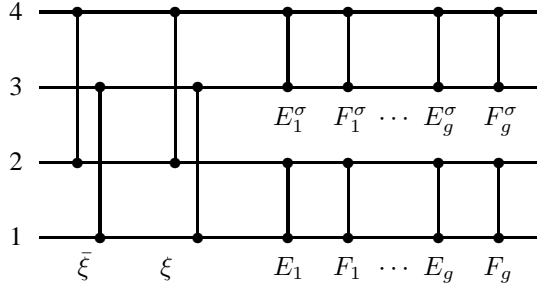
\begin{figure}[htb]
\begin{center}
\unitlength1cm
\begin{picture}(7,4)
\thicklines
\put(0.4,4){\line(1,0){6.6}}
\put(0.4,3){\line(1,0){6.6}}
\put(0.4,2){\line(1,0){6.6}}
\put(0.4,1){\line(1,0){6.6}}
\put(0.9,4){\line(0,-1){2}}
\put(1.2,3){\line(0,-1){2}}
\put(2.2,4){\line(0,-1){2}}
\put(2.5,3){\line(0,-1){2}}
\multiput(3.7,4)(0.8,0){2}{\line(0,-1){1}}
\multiput(3.7,2)(0.8,0){2}{\line(0,-1){1}}
\multiput(5.7,4)(0.8,0){2}{\line(0,-1){1}}
\multiput(5.7,2)(0.8,0){2}{\line(0,-1){1}}
\put(0.9,4){\circle*{0.15}}
\put(0.9,2){\circle*{0.15}}
\put(1.2,3){\circle*{0.15}}
\put(1.2,1){\circle*{0.15}}
\put(2.2,4){\circle*{0.15}}
\put(2.2,2){\circle*{0.15}}
\put(2.5,3){\circle*{0.15}}
\put(2.5,1){\circle*{0.15}}
\multiput(3.7,4)(0.8,0){2}{\circle*{0.15}}
\multiput(3.7,3)(0.8,0){2}{\circle*{0.15}}
\multiput(3.7,2)(0.8,0){2}{\circle*{0.15}}
\multiput(3.7,1)(0.8,0){2}{\circle*{0.15}}
\multiput(5.7,4)(0.8,0){2}{\circle*{0.15}}
\multiput(5.7,3)(0.8,0){2}{\circle*{0.15}}
\multiput(5.7,2)(0.8,0){2}{\circle*{0.15}}
\multiput(5.7,1)(0.8,0){2}{\circle*{0.15}}
\put(0,3.9){4}
\put(0,2.9){3}
\put(0,1.9){2}
\put(0,0.9){1}
\put(0.9,0.5){$\bar{\xi}$}
\put(2,0.5){$\xi$}
\put(3.5,0.5){$E_1$}
\put(4.3,0.5){$F_1$}
\put(4.9,0.5){$\cdots$}
\put(5.5,0.5){$E_g$}
\put(6.3,0.5){$F_g$}
\put(3.5,2.5){$E_1^\sigma$}
\put(4.3,2.5){$F_1^\sigma$}
\put(4.9,2.5){$\cdots$}
\put(5.5,2.5){$E_g^\sigma$}
\put(6.3,2.5){$F_g^\sigma$}
\end{picture}
\end{center}
\caption{The Hurwitz diagram of $\hat{\mathcal{L}}$ shows the Riemann 
surface as seen from the side.}
\label{hurwitz}
\end{figure}

The monodromy matrix $L$ defined in (\ref{eq:monodromy}) corresponding 
to (\ref{a1}) is now also a 
$2\times2$-matrix. Since this matrix can be chosen without loss of 
generality to be tracefree, the characteristic equation 
(\ref{eq:char}) takes the form 
\begin{equation}
    \hat{\mu}^{2}=\mathcal{P}(K)
    \label{eq:muh2},
\end{equation}
where $\mathcal{P}$ is an analytic function in $K$. For polynomial 
$\mathcal{P}=\sum_{i=1}^{g}(K-E_{i})(K-F_{i})$, the two-sheeted 
Riemann surface defined by (\ref{eq:muh2}) is compact. Since the 
spectral parameter $K$ varies on the two-sheeted surface 
$\mathcal{L}$, equation (\ref{eq:muh2}) defines a two-sheeted branched 
cover of $\mathcal{L}$. In other words the so defined Riemann surface 
$\hat{\mathcal{L}}$ is a four-sheeted cover of the complex plane. The 
structure of this surface is shown in the Hurwitz diagram 
Fig.\ \ref{hurwitz}. 

There is an automorphism $\sigma$ of $\hat{\mathcal{L}}$ inherited 
from $\mathcal{L}$
which ensures $E_i^{\sigma}$, $E_i$ and $F_i^{\sigma}$, $F_i$ have 
the same projection in the complex plane. The orbit 
space
$\mathcal{L}_H=\hat{\mathcal{L}}/\sigma$ is then, see~\cite{algebro}, 
again a Riemann
surface given by
\begin{equation}
    \mu^{2}=(K-\xi)(K-\bar{\xi})
\prod_{i=1}^{g}(K-E_{i})(K-F_{i})
    \label{mu}.
\end{equation}
The fixed points $\xi$, $\bar{\xi}$
of the involution $\sigma$ are additional branch 
points of the Riemann surface. The points $E_{i}=\alpha_{i}-i\beta_{i}$
have to be constant with respect to 
the physical coordinates. They are subject to the reality condition 
$E_{i}=\bar{F}_{i}$ or $E_{i},F_{i}\in \mathbb{R}$ for $i=1,\ldots,g$.
A surface of the form (\ref{mu}) is 
called hyperelliptic, since the square root of a polynomial can be 
considered as the straight forward generalization of elliptic 
surfaces. 
Thus it is possible to construct components of the matrix $\Psi$ on 
$\mathcal{L}_H$ which makes it possible to use the powerful 
calculus of hyperelliptic Riemann surfaces.

\subsection{Solutions on hyperelliptic surfaces}
Solutions to the Ernst equation on hyperelliptic surfaces were given 
by Korotkin in \cite{koro88}, and in the gauge (\ref{lin9}) in 
\cite{korot92}. We summarize basic facts of hyperelliptic surfaces 
(see e.g.\ \cite{algebro} and \cite{mumford2} to \cite{fay}) to 
be able to present these solutions.

\begin{figure}[htb]
    \centering 
     \epsfig{file=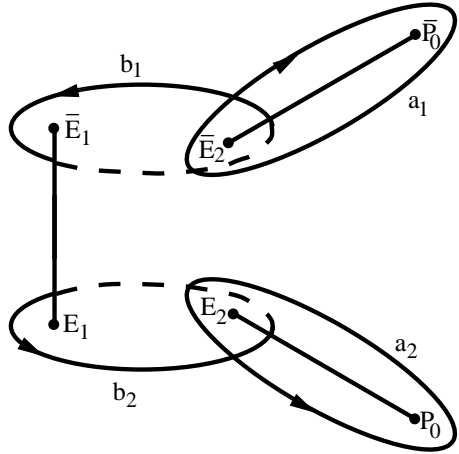,width=6cm}
    \caption{Canonical cycles ($P_{0}=\xi$).}
    \label{fig:cut-system}
\end{figure}

A Riemann surface has a topological invariant, the genus, which 
loosely speaking gives the number of holes in the surface. A surface 
of genus $g>1$ is topologically just a sphere with $g$ handles. For 
genus 0, it is the Riemann sphere, in the elliptic case $g=1$ a 
torus. The hyperelliptic surface $\mathcal{L}_{H}$ defined by (\ref{mu}) 
has genus $g$. We order the branch points with $\mbox{Im}E_{i}<0$ in a way that 
$\Re E_{1}<\Re E_{2}<\ldots <\Re E_{g}$ and assume for 
simplicity that the real parts of the $E_{i}$ are all different, and 
that there are no real branch points.
On this surface we introduce a canonical basis of cycles which 
are non-homologous to zero, i.e.\ which cannot be contracted to a 
point. This basis consists of $2g$ cycles $a_{i}$, $b_{i}$, 
$i=1,\ldots,g$ which do not intersect except for $a_{i}$, $b_{i}$ with 
the same index. For the surface of genus 2 we will consider in detail 
later, we use the cut-system in Fig. \ref{fig:cut-system}. 
The surface resulting from  cutting $\mathcal{L}_{H}$ along these 
cycles, the fundamental polygon,  
is simply connected. 

On a surface of genus $g$, there are $g$ independent holomorphic 
one-forms which are also called differentials of the first kind. 
These one-forms can be locally written as $F(K)dK$ where $F(K)$ is a 
holomorphic function. Their integrals are holomorphic functions. For 
general Riemann surfaces, the determination of the holomorphic differentials 
is a non-trivial problem. One important simplification in the theory 
of hyperelliptic surfaces is that these differentials are explicitly 
known. A basis is provided by 
\begin{equation}
d\nu_k = \left(   \frac{dK}{\mu}, \frac{KdK}{\mu},\ldots,
  \frac{K^{g-1}dK}{\mu} \right)
    \label{basis}.
\end{equation}
The holomorphic differentials
$d\omega_k$ are normalized by the condition on the $a$-periods
\begin{equation}
    \int_{a_{l}}^{}d\omega_{k}=2\pi i \delta_{lk}.
    \label{normholo}
\end{equation}
The matrix of $b$-periods is given by $\mathbf{B}_{ik} =
\int_{b_{i}}^{}d\omega_{k}$. The matrix $\mathbf{B}$ is a so-called
Riemann matrix, i.e.\ it is symmetric and has a negative definite real
part. The Abel map $\omega: \mathcal{L}_{H} \to 
\mbox{Jac}(\mathcal{L}_{H}) $ with base
point $E_{1}$ is defined as $\omega(P)=\int_{E_{1}}^{P}d\omega_k$, 
where $\mbox{Jac}(\mathcal{L}_{H})$ is the Jacobian of $\mathcal{L}_{H}$, 
$\mathbb{C}^{g}$ factorized with respect to the lattice of $a$- and 
$b$-periods.  The
theta function with characteristics corresponding to the curve
$\mathcal{L}_{H}$ is given by
\begin{equation}
    \Theta_{\mathbf{p}\mathbf{q}}(\mathbf{x}|\mathbf{B})=
    \sum_{\mathbf{n}\in\mathbb{Z}^{g}}^{}\exp\left\{\frac{1}{2}
    \langle\mathbf{B}(\mathbf{p}+\mathbf{n}),(\mathbf{p}+\mathbf{n})
    \rangle+\langle\mathbf{p}+\mathbf{n},2\pi i\mathbf{q}+\mathbf{x}
    \rangle\right\}
    \label{theta},
\end{equation}
where $\mathbf{x}\in \mathbb{C}^{g}$ is the argument and
$\mathbf{p},\mathbf{q}\in \mathbb{C}^{g}$ are the characteristics. We
will mainly consider half-integer characteristics in the following. A 
half-integer characteristic is called odd if $4\langle \mathbf{p},
\mathbf{q}\rangle$ is odd,  and even if this inner product is even.  
The theta function with characteristics is, up to an exponential factor,
equivalent to the theta function with zero characteristic (the Riemann
theta function is denoted with $\Theta$) and shifted argument,
\begin{equation}
    \Theta_{\mathbf{p}\mathbf{q}}(\mathbf{x}|\mathbf{B})=
    \Theta(\mathbf{x}+\mathbf{B}\mathbf{p}+2\pi i
    \mathbf{q})\exp\left\{
    \frac{1}{2}\langle\mathbf{B}\mathbf{p},\mathbf{p}
    \rangle+\langle\mathbf{p},2\pi i\mathbf{q}+\mathbf{x} \rangle\right\}.
    \label{theta2}
\end{equation}
The theta function has the periodicity properties 
\begin{equation}
    \Theta_{\mathbf{p}\mathbf{q}}(\mathbf{z}+2\pi i\mathbf{e}_{j}) = 
    e^{2\pi ip_{j}}
    \Theta_{\mathbf{p}\mathbf{q}}(\mathbf{z}),
    \quad 
    \Theta_{\mathbf{p}\mathbf{q}}(\mathbf{z}+\mathbf{B}
    \mathbf{e}_{j})=
    e^{-2\pi i q_{j}-z_{j}-\frac{1}{2}B_{jj}}
    \Theta_{\mathbf{p}\mathbf{q}}(\mathbf{z})
    \label{eq:periodicity},
\end{equation}
where $\mathbf{e}_{j}$ is the $g$-dimensional vector consisting of 
zeros except for a 1 in jth position.

We denote by $d\omega_{PQ}$ a differential of the third kind, i.e., a
one-form which is holomorphic except for two poles in 
$P,Q\in \mathcal{L}_{H}$ with residues
$+1$ and $-1$ respectively. 
This singularity structure characterizes the
differentials only up to holomorphic differentials. They can be
uniquely determined by the normalization
condition that all $a$-periods vanish.  
The differential $d\omega_{\infty^{+}\infty^{-}}$ is
given up to holomorphic differentials by $-K^{g}dK/\mu$. 

In~\cite{prl,prd2} a physically interesting subclass of Korotkin's
solution was identified which can be written in the form
\begin{equation}
    \mathcal{E}=\frac{\Theta_{\mathbf{p}\mathbf{q}}(\omega(\infty^{+})+\mathbf{u})}{
    \Theta_{\mathbf{p}\mathbf{q}}(\omega(\infty^{-})+\mathbf{u})}
  e^{I}
    \label{ernst2},
\end{equation}
where the characteristic is subject to the reality condition 
$\mathbf{B}\mathbf{p}+\mathbf{q}\in i\mathbb{R}$,
where $\mathbf{u}=(u_k)\in\mathbb{C}^g$ and where
\begin{equation}
    I=\frac{1}{2\pi i}\int_{\Gamma}^{}\ln G(K)\,d\omega_{\infty^{+}
    \infty^{-}}(K), \qquad u_k=\frac{1}{2\pi i}
    \int_{\Gamma}^{}\ln G(K)\,d\omega_k.
    \label{path1}
\end{equation}
$\Gamma$ is a piece-wise smooth contour on $\mathcal{L}_{H}$ and $G(K)$ is
a non-zero H\"older-continuous function on $\Gamma$. The contour
$\Gamma$ and the function $G$ have to satisfy the reality conditions
that with $K\in \Gamma$ also $\bar{K}\in \Gamma$ and
$\bar{G}(\bar{K})=G(K)$; both are independent of the physical
coordinates. In the case of disks of radius 1 in which we are 
interested here, the contour $\Gamma$ is the
covering of the imaginary axis in the +-sheet of $\mathcal{L}_{H}$ between
$-i$ and $i$.
In \cite{prd2} it was 
shown that solutions of the above form on a Riemann surface of even 
genus $g=2s$ given by 
$\mu^{2}=(K-\xi)(K-\bar{\xi})\prod_{i=1}^{s}(K^{2}-E_{i}^{2})(K^{2}-
\bar{E}_{i}^{2})$ with a function $G$ subject to
$G(-K)=\bar{G}(K)$ lead to an equatorially symmetric Ernst potential, 
$\mathcal{E}(\zeta)=\bar{\mathcal{E}}(-\zeta)$.

Notice that these solutions depend only via the branch points of 
$\mathcal{L}_{H}$ on the physical coordinates.  For $g=0$ there are no 
theta functions and the potential (\ref{ernst2}) takes the form
\begin{equation}
    \ln \mathcal{E}=\frac{1}{2\pi i}\int_{\Gamma}^{}\frac{\ln GdK}{
    \sqrt{(K-\zeta)^{2}+\rho^{2}}}
    \label{eq:newtona}.
\end{equation}
As shown in section 2, this Ernst potential is equivalent 
to the Poisson integral with a distributional density at the disk. 
The solutions can be considered as the static or 
Newtonian limit of the theta functional solutions (\ref{ernst2}).  

In the theta functional solutions of evolution equations, the 
underlying Riemann surface is independent of the physical 
coordinates. The coordinates only occur in the argument of the theta 
functions. Because of (\ref{eq:periodicity}) 
the solutions are periodic or quasi-periodic. In 
the case of the Ernst equation, the Riemann surface itself is 
`dynamical' since some of its branch points are parametrized by the 
physical coordinates. Thus the solution is given on a whole 
family $\mathcal{L}_{H}(\xi,\bar{\xi})$ of surfaces. The argument of 
the theta functions in (\ref{ernst2}) has however no additional 
dependence on the physical coordinates. Here the modular dependence of 
the theta functions is important. Therefore the solutions show no 
periodicity and can be asymptotically flat. 

The limit $E_{i}\to F_{i}$ in which the cut collapses corresponds 
to the solitonic limit for theta functional solutions to nonlinear evolution 
equations. The almost periodic solutions can be seen as an infinite 
train of solitons. In the `solitonic' limit, only a finite number of 
solitons survive. In the case of the Ernst equation, the Kerr 
solution can be obtained as the corresponding limit of a genus 2 
surface, see \cite{koro88}. In this mathematical sense, black holes 
can be considered as solitons. The defining equation (\ref{mu}) for 
the Riemann surface illustrates the relation between solitons and 
theta functional solutions: the solitons have a different monodromy 
property which can be obtained in the limit of a degenerate surface 
(collapsing cut). Zeros of first order of $\mathcal{P}$ in 
(\ref{eq:muh2}) correspond to 
theta functional solutions, zeros of order 2 to solitons. 
The $N$-solitons \cite{belzak} and the B\"acklund 
transformations are thus contained in Korotkin's solutions as a 
limiting case. 

The importance of many exact solutions to Einstein's equations (see 
\cite{exac}) is 
somewhat limited by the singularities they can have. In the stationary 
vacuum, the Lichnerowicz theorem \cite{lichner} states that a solution 
is either Minkowski spacetime or it has singularities. Non-trivial 
solutions must have sources which show in the form of singularities 
in matterfree settings. For general explicit solutions to the Einstein 
equations, it is however difficult to localize the singularities and 
to show that they can be replaced by matter sources. This is also in 
general true for Korotkin's solutions. 
An important advantage of the subclass discussed here is 
that general statements can be made on the regularity of the solutions:
\begin{theorem}
    The Ernst potential (\ref{ernst2}) with  $[\mathbf{p}\mathbf{q}]$ 
    being a non-singular half-integer characteristic is analytic in 
    the exterior of the disk $\zeta=0$, $0\leq \rho\leq 1$ where it is 
    discontinuous iff
    \begin{equation}
	\Theta_{\mathbf{p}\mathbf{q}}(\omega(\infty^{-})+\mathbf{u}) 
	\neq 0
        \label{eq:condzero}.
    \end{equation}
\end{theorem}
For a proof see \cite{prd2}. Condition (\ref{eq:condzero}) 
reflects the fact that within general relativity arbitrary amounts of 
energy cannot be concentrated in a finite region of spacetime without 
forming a black hole or a singularity. This defines the 
range of the physical parameters where the solution is regular. A 
general method how to define this parameter range 
based on a study of the zeroes of theta functions was developed in 
\cite{prd3}.

The  Ernst potential (\ref{ernst2})
follows with (\ref{lin9}) and $J=\frac{1}{2\pi i}\int_{\Gamma}^{}\ln
G d
\omega_{PP^{\sigma}}$ from a matrix $\Psi$ of the form
\begin{equation}\label{a7}
    \Psi=e^{I/2}\sqrt{\frac{\det(\infty)}{\det(K)}}
    \left(
    \begin{array}{cc}
	\frac{\Theta_{\mathbf{p}\mathbf{q}}(\mathbf{u}+\omega(P))}{
	\Theta_{\mathbf{p}\mathbf{q}}(\mathbf{u}+\omega(\infty^{-}))}e^{J/2}
     & -i 
    \frac{\Theta_{\mathbf{p}\mathbf{q}}(\mathbf{u}+\omega(P^{\sigma}))}{
    \Theta_{\mathbf{p}\mathbf{q}}(\mathbf{u}+\omega(\infty^{-}))}e^{-J/2}
     \\
    \frac{\Theta_{\mathbf{p}\mathbf{q}}(\mathbf{u}+\omega(P)+
	\omega(\bar{\xi}))}{
	\Theta_{\mathbf{p}\mathbf{q}}(\mathbf{u}+\omega(\infty^{-})+\omega(\bar{\xi}))}e^{
	J/2}
	& i
   \frac{ \Theta_{\mathbf{p}\mathbf{q}}(\mathbf{u}+\omega(P^{\sigma})+\omega(\bar{\xi}))
       }{
       \Theta_{\mathbf{p}\mathbf{q}}(\mathbf{u}+\omega(\infty^{-})+\omega(\bar{\xi}))}e^{
       -J/2}
    \end{array}\right),
\end{equation}
with $\det(K) =\Theta_{\mathbf{p}\mathbf{q}}(\mathbf{u}+\omega(P))
\Theta_{\mathbf{p}\mathbf{q}}(\mathbf{u}-\omega(P)+
    \omega(\bar{\xi}))+\Theta_{\mathbf{p}\mathbf{q}}(\mathbf{u}-\omega(P))
    \Theta_{\mathbf{p}\mathbf{q}}(\mathbf{u}+\omega(P)+
    \omega(\bar{\xi}))$.

To determine the metric function $a$ via (\ref{a5}), we have to 
calculate the matrix 
$\mathcal{S}$ which leads with 
(\ref{a7}) to 
\begin{equation}
    \mathcal{S}_{12}=\frac{1}{2if}D_{\infty^{+}}\ln\frac{\Theta_{\mathbf{p}\mathbf{q}}
    (\mathbf{u}+\omega(\infty^{-}))
    }{\Theta_{\mathbf{p}\mathbf{q}}(\mathbf{u}+\omega(\infty^{-})+\omega(\bar{\xi}))}
    \label{12},
\end{equation}
and to determine the Harrison transformed function $a$ in the next 
section,
\begin{equation}
    \mathcal{S}_{21}=-\frac{\mathcal{E}\bar{\mathcal{E}}}{2if}
    D_{\infty^{+}}\ln\frac{\Theta_{\mathbf{p}\mathbf{q}}
    (\mathbf{u}+\omega(\infty^{+}))
    }{\Theta_{\mathbf{p}\mathbf{q}}(\mathbf{u}+\omega(\infty^{+})+\omega(\bar{\xi}))}
    \label{21}.
\end{equation}
Using a degenerated version of Fay's trisecant identity \cite{fay} (see 
\cite{prd2} for the present case), we can write the above relations 
free of derivatives,
\begin{eqnarray}
    \mathcal{S}_{12}&=&\frac{\rho}{2f}\left(\frac{\Theta_{\mathbf{p}\mathbf{q}}
    (\mathbf{u})\Theta_{\mathbf{p}\mathbf{q}}(\mathbf{u}+
    2\omega(\infty^{-})+\omega(\bar{\xi}))}{L\Theta_{\mathbf{p}\mathbf{q}}(\mathbf{u}+
    \omega(\infty^{-})+\omega(\bar{\xi}))\Theta_{\mathbf{p}\mathbf{q}}(\mathbf{u}+
    \omega(\infty^{-}))}-1\right)\nonumber\\
    \mathcal{S}_{21}&=&-\frac{\rho\mathcal{E}\bar{\mathcal{E}}}{2f}
    \left(\frac{\Theta_{\mathbf{p}\mathbf{q}}(\mathbf{u})
    \Theta_{\mathbf{p}\mathbf{q}}(\mathbf{u}+
    2\omega(\infty^{+})+\omega(\bar{\xi}))}{L\Theta_{\mathbf{p}\mathbf{q}}(\mathbf{u}+
    \omega(\infty^{+})+\omega(\bar{\xi}))\Theta_{\mathbf{p}\mathbf{q}}(\mathbf{u}+
    \omega(\infty^{+}))}-1\right)
    \label{Z'},
\end{eqnarray}
where 
\begin{equation}
    L=\frac{\Theta(\omega(\infty^{-}))\Theta(\omega(\infty^{-})+
    \omega(\bar{\xi}))}{\Theta(0)\Theta(
    \omega(\bar{\xi}))}
    \label{C}.
\end{equation}

To construct the solution for the counter-rotating disks in \cite{prd3}, 
we used an algebraic approach which made it possible to establish 
algebraic relations between the metric functions at the disk. 
Let us recall that a divisor $X$ on $\mathcal{L}_{H}$
is a formal symbol  $X=n_{1}P_{1}+\ldots+ n_{k}P_{k}$ with $P_{i}\in 
\mathcal{L}_{H}$ and $n_{i}\in \mbox{Z}$. The degree of a divisor is 
$\sum_{i=1}^{k}n_{i}$.  The Riemann vector $K_{R}$ is 
defined by the condition that $\Theta(\omega(W)+K_{R})=0$ if $W$ is a 
divisor of degree $g-1$ or less. We use here and in the following the 
notation $\omega(W)=\int_{P_{0}}^{W}\mathrm{d}\omega
=\sum_{i=1}^{g-1}\omega(W_{i})$. 
Note that  the Riemann vector can be 
expressed through half-periods in the case of a hyperelliptic surface.
We define the divisor $X=\sum_{i=1}^{g}X_{i}$ 
as the solution of the Jacobi inversion problem ($i=1,\ldots,g$)
\begin{equation}
	\omega(X)-\omega(D)=\mathbf{u}
	\label{eq18},
\end{equation}
where the divisor $D=\sum_{i=1}^{g}E_{i}$ (this corresponds to a 
choice of the characteristic  in (\ref{ernst2})). 
With the help of these divisors, we can write (\ref{ernst2}) in the form
\begin{equation}
	\ln \mathcal{E}=\int_{D}^{X}\frac{\tau^{g}d\tau}{
	\mu(\tau)}-\frac{1}{2\pi \mathrm{i}}
	\int_{\Gamma}^{}\ln G \frac{\tau^{g}d\tau}{\mu(\tau)}
	\label{eq19}.
\end{equation}
Additional information follows from the reality of $u$ 
which leads to $\omega(X)-\omega(D)=\omega(\bar{X})-\omega(\bar{D})$. 
The reality condition for $X$ implies via Abel's theorem the existence 
of a meromorphic function $R$ with poles in $\bar{X}+D$ and zeros in 
$X+\bar{D}$ (which is a rational function in the fundamental polygon),
\begin{equation}
    R(K)=const \frac{\prod_{i=1}^{g}(K-E_{i})(K-\bar{\mathcal{E}}_{i})-Q_{0}(K)\mu(K)}{
    \prod_{i=1}^{g}(K-\bar{X}_{i})(K-E_{i})}
    \label{abel},
\end{equation}
where $Q_{0}(K)=x_{0}+x_{1}K+\ldots+xK^{g-1}$ is a polynomial in 
$K$ with purely imaginary coefficients and $x=ibe^{-2U}$. 
The coefficients $x_{i}$ are related 
to $X$ via the relation 
\begin{equation}
    (1-x^{2})\prod_{i=1}^{g}(K-X_{i})(K-\bar{X}_{i})=
    \prod_{i=1}^{g}(K-E_{i})(K-\bar{\mathcal{E}}_{i})-Q_{0}^{2}(K)(K-\xi)(K-\bar{\xi})
    \label{abela}.
\end{equation}
We can use the existence of the 
rational function $R$ to calculate 
certain integrals of the third kind as 
\begin{equation}
    \frac{\Theta(u+\omega(P))\Theta(u+\omega(P^{\sigma})+
\omega(\bar{\xi})) 
}{\Theta(u+\omega(P^{\sigma}))\Theta(u+\omega(P)+\omega(\bar{\xi}))}
=\exp\left(\int_{X+\bar{D}}^{\bar{X}+D}d\omega_{P^{\sigma}P}
\right)=\frac{R(P)}{R(P^{\sigma})}
    \label{abel2}.
\end{equation}
This makes it possible to give an algebraic expression for $S_{12}$ 
and $S_{21}$. We can write $S_{12}$ in the form
\begin{equation}
    \mathcal{S}_{12}=\frac{1}{i(\mathcal{E}+\bar{\mathcal{E}})}
    D_{\infty^{+}}\left(\int_{X}^{\bar{X}}d\omega_{PQ}
    \right)=\frac{1}{i(\mathcal{E}+\bar{\mathcal{E}})}
    D_{\infty^{+}}\left(\int_{X+\bar{D}}^{\bar{X}+D}d\omega_{PQ}
    +\int_{D}^{\bar{D}}d\omega_{PQ}
    \right)
    \label{12a}
\end{equation}
with $Q$ independent of $P$. 
The second integral can be reexpressed in terms of theta functions and 
be calculated with the help of so-called root 
functions:
The quotient of two theta functions with the same argument but 
different characteristic is a  root function which means that 
its square is a function on $\mathcal{L}_{H}$. 
Let  $P_{i}$, $i=1,\ldots, 2g+2$, be 
the branch points of a hyperelliptic Riemann surface $\mathcal{L}_{H}$ of genus $g$ 
and $A_{j}=\omega(P_{j})$ with $\omega(P_{1})=0$. Furthermore let 
$\{i_{1},\ldots, i_{g}\}$ and $\{j_{1},\ldots,j_{g}\}$ be two sets of 
numbers in $\{1,2,\ldots,2g+2\}$. Then the following 
equality holds for an 
arbitrary point $P\in \mathcal{L}_{H}$,
\begin{equation}
	\frac{\Theta\left[K_{R}+\sum_{k=1}^{g}A_{i_{k}}\right]\left(
	\omega(P)\right)}{\Theta\left[K_{R}+
	\sum_{k=1}^{g}A_{j_{k}}\right]\left(
	\omega(P)\right)}=c_{1}\sqrt{\frac{(K-E_{i_{1}})\ldots(K-E_{i_{g}})}{
	(K-E_{j_{1}})\ldots(K-E_{j_{g}})}}
	\label{root1},
\end{equation}
where $c_{1}$ is a constant independent of $K$.

This implies 
\begin{equation}
    \exp\left(\int_{D}^{\bar{D}}d\omega_{P^{\sigma}\infty^{+}}\right)
= \pm \prod_{i=1}^{g}\sqrt{\frac{K-F_{i}}{K-E_{i}}}
    \label{root2}.
\end{equation}
Thus we get with (\ref{abel2})
\begin{equation}
    \mathcal{S}_{12}=\frac{1}{i(\mathcal{E}+\bar{\mathcal{E}})}\left(\frac{1}{2}
    \sum_{i=1}^{g}(X_{i}-\bar{X}_{i})+\frac{1}{1-x^{2}}
    \left(\frac{x}{2}\left(\sum_{i=1}^{g}(E_{i}+\bar{E}_{i})-(\xi+\bar{\xi})
    \right)+x_{g-2}\right)\right),
    \label{12b}
\end{equation}
and similarly
\begin{equation}
    \mathcal{S}_{21}=-\frac{\mathcal{E}\bar{\mathcal{E}}}{i(\mathcal{E}+
    \bar{\mathcal{E}})}\left(\frac{1}{2}
    \sum_{i=1}^{g}(X_{i}-\bar{X}_{i})-\frac{1}{1-x^{2}}
    \left(\frac{x}{2}\left(\sum_{i=1}^{g}(E_{i}+\bar{E}_{i})-(\xi+\bar{\xi})
    \right)+x_{g-2}\right)\right).
    \label{21b}
\end{equation}

\subsection{Counter-rotating dust disk}
As an example we will discuss a class of disk solutions \cite{prl2,prd3}
on a genus 2 surface where the disk can be interpreted as 
two
counter-rotating components of pressureless matter, so-called dust. 
The surface energy-momentum tensor $S^{\alpha\beta}$ 
of these models, where $\alpha$ and $\beta$ 
stand for the $t$, $\rho$ and $\phi$ components, 
is defined on the
hypersurface $\zeta=0$.  The tensor $S^{\alpha\beta}$ is related to the 
energy-momentum tensor $T^{\alpha\beta}$ which appears in the Einstein 
equations $G^{\mu\nu}=8\pi T^{\mu\nu}$ 
via $T^{\alpha\beta}=S^{\alpha\beta}e^{k-U}\delta(\zeta)$. 
The tensor $S^{\alpha\beta}$ can be written in the form
\begin{equation} 
    S^{\alpha\beta}=\sigma_+ u^{\alpha}_+ u^{\beta}_+ +\sigma_{-}
u^{\alpha}_- u^{\beta}_- \label{2.0},
\end{equation}
where $u_{\pm}=(1,0,\pm \Omega)$.  We gave an explicit solution
for disks with constant angular velocity $\Omega$ and constant
relative density
$\gamma=(\sigma_{+}-\sigma_{-})/(\sigma_{+}+\sigma_{-})$.
A physical interpretation of the surface energy-momentum tensor 
$S^{\alpha\beta}$ will be given in the following section. 

This class
of solutions is characterized by two real parameters $\lambda$ and
$\delta$ which are related to $\Omega$ and $\gamma$ and the metric
potential $U_{0}=U(0,0)$ at the center of the disk via
\begin{equation} \label{2.1}
\lambda=2\Omega^{2}e^{-2U_{0}}, \quad 
\delta=\frac{1-\gamma^{2}}{\Omega^{2}}.
\end{equation}
We put the radius $\rho_{0}$ of the disk equal to 1 unless otherwise
noted.  Since the radius appears only in the combinations
$\rho/\rho_{0}$, $\zeta/\rho_{0}$ and $\Omega \rho_{0}$ in the physical
quantities, it does not have an independent role.  It is always
possible to use it as a natural length scale unless
it tends to 0 as in the case of the ultrarelativistic limit of the one
component disk.  The Ernst potential
will be discussed in dependence of
the parameters $\epsilon=z_{R}/(1+z_{R})=1-e^{U_{0}}$  and $\gamma$, 
where $z_{R}$ is the redshift of photons emitted at the center of the 
disk and detected at infinity.

The solution is given on a surface of genus 2 where
the branch points of the Riemann surface are given by the relation 
$E_{1}=-\bar{E}_{2}$ and
$E:=E_{1}^{2}=\alpha+\mathrm{i}\beta$ where
\begin{equation} \alpha=-1+\frac{\delta}{2},\quad
\beta=\sqrt{\frac{1}{\lambda^{2}}+\delta -\frac{\delta^{2}}{4}}
\label{2.3}.
\end{equation}
The function $G$ in (\ref{path1}) reads
\begin{equation}
G(\tau)=\frac{\sqrt{(\tau^{2}-\alpha)^2+\beta^2}+\tau^{2}+1}{
\sqrt{(\tau^{2}-\alpha)^2+\beta^2}-(\tau^{2}+1)}.  \label{2.4}
\end{equation}
We note that with $\alpha$ and $\beta$ given, the
Riemann surface is completely determined at a given point in the
spacetime, i.e.\ for a given value of $\xi$. 

Regularity of the solutions in the exterior of the disk
restricts the physical parameters to
$0\leq \delta\leq
\delta_{s}(\lambda):=2\left(1+\sqrt{1+1/\lambda^{2}}\right)$ and $0<\lambda
\leq \lambda_{c}$ where $\lambda_{c}(\gamma)$ is the smallest value of
$\lambda$ for which $\epsilon=1$.  
The range of the physical parameters is restricted by the following 
limiting cases:\\
\emph{Newtonian limit}: $\epsilon=0$ ($\lambda=0$), i.e.\ small 
velocities $\Omega \rho_{0}$ and small redshifts in the disk. 
The function $e^{2U}$ tends independently of $\gamma$  
to $1+\lambda U_{N}$, where $U_{N}$ is the Maclaurin disk 
solution, and $b$ is of order 
$\Omega^{3}$. \\
\emph{Ultrarelativistic limit}: $\epsilon=1$, i.e.\ 
diverging central redshift. For $\gamma\neq1$ it is reached for 
$\lambda_{c}=\infty$. The solution describes a disk of finite 
extension with diverging 
central redshift.
For $\gamma=1$, the limit is reached for  $\lambda_{c}=4.629\ldots$. 
In this case the solution has a singular axis and is not 
asymptotically flat. This behavior can be interpreted as the limit 
of a vanishing disk radius. With this rescaling the solution in the 
exterior of the disk can be interpreted as the extreme Kerr solution 
(see \cite{prd4} and references given therein).
\\
\emph{Static limit}: $\gamma=0$ ($\delta=\delta_{s}(\lambda)$). 
In this limit,  the solution belongs to the Morgan and Morgan class 
\cite{morgan}. 
\\
\emph{One component}: $\gamma=1$ ($\delta=0$), i.e.\ no counter-rotating 
matter in the disk. This is the disk of \cite{bawa,neugebauermeinel}.

Analytic formulas for the complete metric in terms of theta functions 
are given in \cite{prd4}. To evaluate the hyperelliptic integrals in 
the expressions for the metric we use the numerical methods
of \cite{prd4}.

At the disk the branch points $\xi,\bar{\xi}$ lie on the contour $\Gamma$ 
which 
implies that care has to be taken in the evaluation of the line integrals. 
The situation is however simplified by the equatorial symmetry of 
the solution which is reflected by the additional involution $K\to -K$
of the Riemann surface $\Sigma_{2}$ for $\zeta=0$. 
This makes it possible to perform the reduction $K^{2}\to \tau$ and 
to express the metric 
in terms of elliptic theta functions (see \cite{prd2}). 
We denote with $\Sigma_{w}$ the elliptic Riemann surface defined by
$\mu_{w}^{2}=(\tau+\rho^{2})((\tau-\alpha)^{2}+\beta^{2})$, and let $dw$
be the associated differential of the first kind with
$u_{w}=\frac{1}{i\pi}\int_{-\rho^{2}}^{-1}\ln G(\sqrt{\tau})dw(\tau)$. 
 We cut the 
surface in a way that the $a$-cut is a closed contour in the upper 
sheet around the cut $[-\rho^{2},\bar{E}]$ and that the $b$-cut starts 
at the cut $[\infty,E]$. The Abel map $w$ is defined for $P\in 
\Sigma_{w}$ as $w(P)=\int_{\infty}^{P}dw$.
Then the real part of the Ernst potential at the disk can be written as
\begin{eqnarray}
e^{2U}&=&\frac{1}{Y-\delta}\left(-\frac{1}{\lambda}-\frac{Y}{\delta}
\left(\frac{\frac{1}{\lambda^{2}}+\delta}{
\sqrt{\frac{1}{\lambda^{2}}+\delta\rho^{2}}}-\frac{1}{\lambda}\right)
\right.  \nonumber\\ &&\left. 
+\sqrt{\frac{Y^{2}((\rho^{2}+\alpha)^{2}+\beta^{2})}{
\frac{1}{\lambda^{2}}+\delta\rho^{2}}-2Y(\rho^{2}+\alpha) +
\frac{1}{\lambda^{2}}+\delta\rho^{2}}\right) \label{y1},
\end{eqnarray}
where
\begin{equation}
Y=\frac{\frac{1}{\lambda^{2}}+\delta\rho^{2}}{\sqrt{(\rho^{2}+\alpha)^{2}
+\beta^{2}}}\frac{\vartheta_{3}^{2}(u_{w})}{\vartheta_{1}^{2}(u_{w})}
\label{y2}.
\end{equation}
It was shown that there exist algebraic relations between the
real and imaginary parts of the Ernst potential, 
\begin{equation}
\frac{\delta^{2}}{2}(e^{4U}+b^{2})= \left(\frac{1}{\lambda}-\delta
e^{2U}\right)\left(\frac{\frac{1}{\lambda^{2}
}+\delta}{\sqrt{\frac{1}{\lambda^{2}}+\delta \rho^{2}}}-
\frac{1}{\lambda}\right)+\delta\left(\frac{\delta+\rho^{2}}{2}-1\right)
\label{2.9},
\end{equation}
and the function $Z:=(a-a_{0})e^{2U}$: 
\begin{equation}
   Z^{2}-\rho^{2}+\delta e^{4U}=\frac{2}{\lambda}e^{2U} \label{2.10a}.
\end{equation}
Moreover we have
\begin{equation}
    ix_{0}=-\frac{Z}{\delta f}\left(\frac{1/\lambda^{2}+
    \delta}{\sqrt{1/\lambda^{2}+\delta \rho^{2}}}-\frac{1}{\lambda}
    \right)
    \label{x0}.
\end{equation}

At the disk, the normal derivatives of the 
metric functions are discontinuous, but they 
can be expressed in terms of 
$\rho$-derivatives via
\begin{eqnarray}
    (e^{2U})_{\zeta} & = & \frac{Z^{2}+\rho^{2}+\delta e^{4U}}{2Z\rho}b_{\rho}
    \nonumber,  \\
    b_{\zeta} & = & -\frac{Z^{2}+\rho^{2}+\delta e^{4U}}{2Z\rho}
    (e^{2U})_{\rho}+\frac{e^{2U}}{Z}
    \label{diskz}.
\end{eqnarray}
This makes it directly 
possible to determine all quantities in the disk in 
terms of elliptic functions.

\subsection{Metric and Harrison transformation}
If stationary axisymmetric solutions for the pure vacuum are 
submitted to a Harrison transformation, the complete transformed metric 
can be constructed. 
The metric function $k$ is invariant under the action of $SU(2,1)$ 
transformations. 
To determine the transformed metric function $a'$, we 
consider the matrix $\mathcal{S}$ in (\ref{12}). If we go over from 
$2\times 2$-matrices to 
$3\times 3$-matrices according to the rule
\begin{equation}
    \left(
    \begin{array}{cc}
	A_{11} &A_{12}   \\
	A_{21} & A_{22}
    \end{array}\right)\to  \left(
    \begin{array}{ccc}
	A_{11} & 0 & A_{12}  \\
	0 & 1 & 0 \\
	A_{21} & 0 & A_{22}
    \end{array}\right)
    \label{rule},
\end{equation}
the matrix $\mathcal{G}$ acts on  $S$ as on $\chi$.
Thus we get with (\ref{group10}) 
\begin{equation}
    \mathcal{S}_{12}'=\frac{1}{(1-q^{2})^{2}}(\mathcal{S}_{12}+
    2iq^{2}-q^{4}\mathcal{S}_{21})
    \label{ehlers4},
\end{equation}
which is in accordance with (\ref{a}). This implies with (\ref{a4})
for the function $a'$
\begin{equation}
    a'-a_{0}'=-\frac{2}{(1-q^{2})^{2}}(\mathcal{S}_{12}-q^{4}
    \mathcal{S}_{21})
    \label{a'}.
\end{equation}
To determine $a_{0}'$, one has to consider $\mathcal{S}_{12}$ and $
\mathcal{S}_{21}$ on 
the axis. In the limit $\rho\to0$, there is a non-trivial contribution 
from the quotient of theta functions in (\ref{Z'}) which diverges as 
$1/\rho$. Repeating the considerations of \cite{prd2} in the 
calculation of $a_{0}$, one finds that the axis potentials can be 
expressed in terms of theta functions on the surface $\tilde{\Sigma}$ 
of genus $g-1$ given by $\tilde{\mu}^{2}=\prod_{i=1}^{g}(K-E_{i})(K-
\bar{E}_{i})$. Denoting quantities on $\tilde{\Sigma}$ with a tilde, 
one has with \cite{prd2} on the axis
\begin{equation}
    \frac{\mathcal{S}_{21}}{\mathcal{S}_{12}}=
    \frac{\tilde{\Theta}_{\tilde{\mathbf{p}}\tilde{\mathbf{q}}}(\tilde{\mathbf{u}}+
	2\tilde{\omega}(\infty^{-}))}{\tilde{\Theta}_{\tilde{\mathbf{p}}\tilde{\mathbf{q}}}(\tilde{\mathbf{u}}+
	2\tilde{\omega}(\infty^{+}))}
    \label{a0a}.
\end{equation}
In the equatorially symmetric case, this quotient is identical to one 
since $2\tilde{\omega}(\infty^{+})$ is a half period on 
$\tilde{\Sigma}$ (see \cite{prd2}). Thus we have
\begin{equation}
    a_{0}'=a_{0}\frac{1+q^{4}}{(1-q^{2})^{2}}
    \label{Z0}.
\end{equation}

To illustrate the class of Harrison transformed hyperelliptic 
solutions, we will now study the transformed counter-rotating dust 
disks. The metric function 
$f'$ in (\ref{f'}) is proportional to $f$, which implies that
the transformed solution vanishes exactly where the original solution 
has zeros. Since the set of 
zeros of $f$ just defines
the ergoregions, the transformed solution has the same ergoregions 
(if any) as the original solution. For the ergoregions of the 
counter-rotating dust disks see \cite{prd4}. 

\begin{figure}[htb]
    \centering
     \epsfig{file=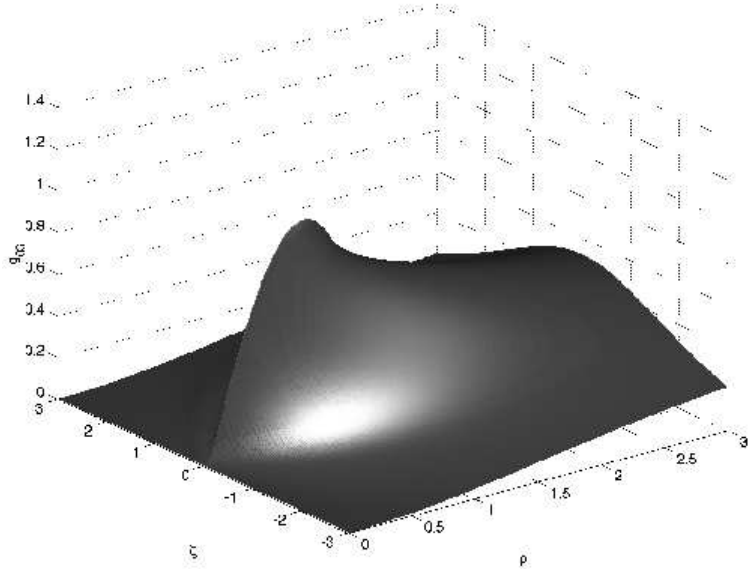,width=10cm}
    \caption{Metric function $g_{03}'$ for $\epsilon=0.85$, 
    $\gamma=0.95$ and $q=0.6$.}
    \label{Aprime}
\end{figure}

For small $q$, the functions $f$ and $b$ are essentially unchanged 
since they are quadratic in $q$. The electromagnetic potential $\Phi$ 
is in this limit with (\ref{Phi'}) of the form
\begin{equation}
    \Phi'=-q(1-f+ib)
    \label{qs}.
\end{equation}
For larger $q$, $|f'|$  becomes smaller near the origin. Since 
its asymptotic values are not changed, the growth  
rate towards infinity increases which is reflected by the mass 
formula (\ref{group12}). In the singular 
limit $q\to 1$, the function $f'$ is 
zero for all finite values of $z$, but one at infinity. The behavior 
for $b'$ is similar with the exception that $b'$ is odd and 
zero at infinity. 
The function $a$ also becomes singular in the limit $q\to 1$ 
which is reflected by the diverging factor $1/(1-q^{2})^{2}$ and the 
constant $a_{0}$ (\ref{Z0}) 
which just implies that one can no longer choose $a$ to be 
zero on the axis. In the metric function $g_{03}'=-a'f'$, the 
factors $(1-q^{2})^{2}$ just cancel and the function is only 
marginally changed with increasing $q$. The typical behavior of 
$g_{03}'$ for values of 
$q$ with $0<|q|<1$ can be seen in Fig. \ref{Aprime}.
The metric function is an even function in $\zeta$ which vanishes on 
the axis and at infinity. It is analytic except at the disk where the 
normal derivatives have a jump.

\begin{figure}[htb]
    \centering
     \epsfig{file=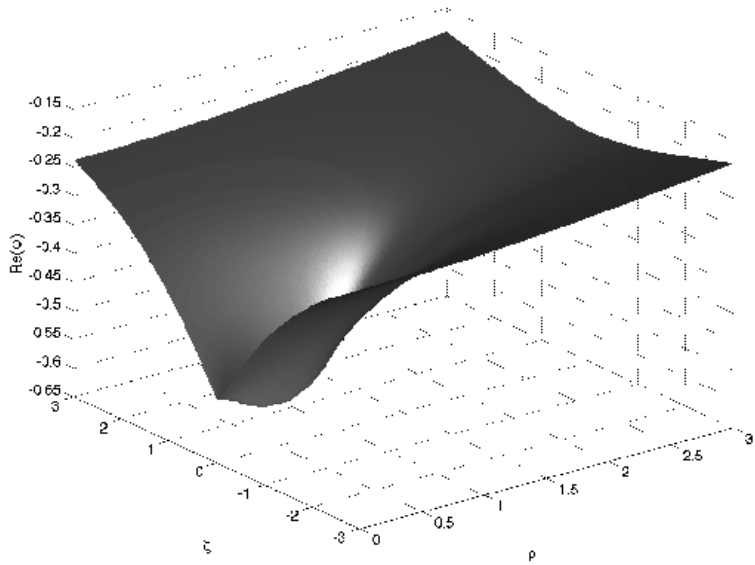,width=10cm}
    \caption{Real part of the electromagnetic potential $\Phi$ 
    for $\epsilon=0.85$, 
    $\gamma=0.95$ and $q=0.6$.}
    \label{RPhi'}
\end{figure}

\begin{figure}[htb]
    \centering
     \epsfig{file=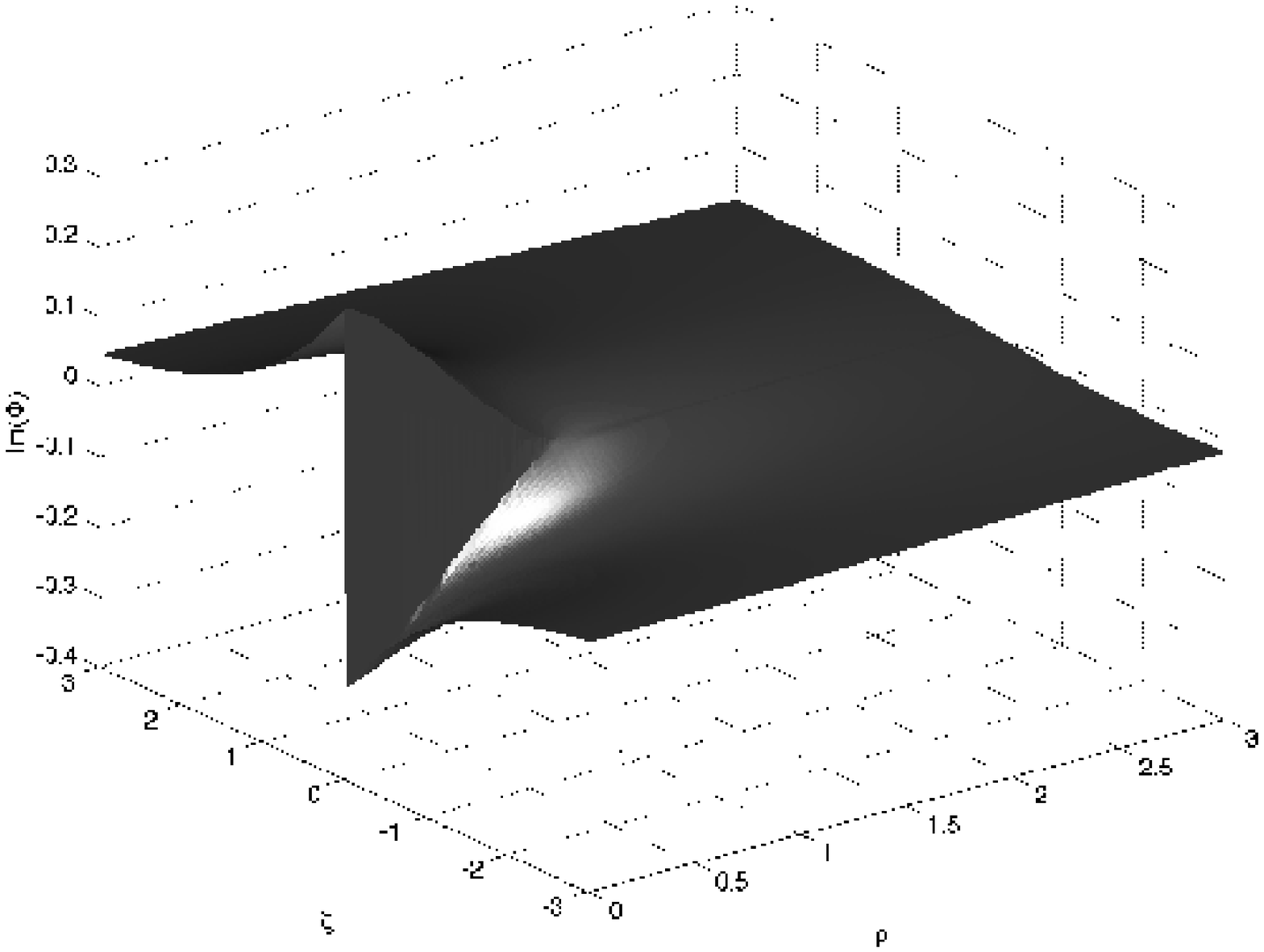,width=10cm}
    \caption{Imaginary part of the electromagnetic potential $\Phi$ 
    for $\epsilon=0.85$, 
    $\gamma=0.95$ and $q=0.6$.}
    \label{IPhi'}
\end{figure}

The electromagnetic potential tends to $-1$ in the limit $q\to 1$ for 
finite $\xi$, but is zero at infinity. The imaginary part is directly 
proportional to $b'$ as can be seen from (\ref{b'}) and (\ref{Phi'}).
We show a typical situation 
for values of $q$ with $0<|q|<1$ in Fig. \ref{RPhi'} for the real 
part and 
in Fig. \ref{IPhi'} for the imaginary part.
The real part of $\Phi$ is an even function in $\zeta$ which vanishes 
at infinity and has discontinuous normal derivatives at the disk.
The imaginary part of $\Phi$ is an odd function in $\zeta$ and has a 
jump at the disk. It vanishes at infinity.

Since $f'$ has the same zeros as $f$, the transformed solution has a 
diverging central redshift if the untransformed has, i.e.\ the 
ultrarelativistic limits coincide. In the case $\gamma\neq1$, one has 
a charged disk of finite extension 
with diverging central redshift. For $\gamma=1$, the 
solution in the exterior of the disk can be interpreted as 
an extreme Kerr-Newman metric which 
is obtained as a Harrison-transformed extreme Kerr metric. 

In the Newtonian limit $\lambda\to 0$, one has $f=1+\lambda U_{N}$ 
and $b=\lambda^{\frac{3}{2}}\tilde{b}$ in lowest order. This implies 
with (\ref{f'}) to (\ref{Phi'}) for $1-q^{2}\gg\lambda$
\begin{equation}
    f'=1+\frac{1+q^{2}}{1-q^{2}}\lambda U_{N},\quad 
    b'=\frac{1+q^{2}}{1-q^{2}}\lambda^{\frac{3}{2}}\tilde{b},\quad
    \Phi'=-\frac{q}{1-q^{2}}(\lambda U_{N}+i
    \lambda^{\frac{3}{2}}\tilde{b}),
    \label{newt}
\end{equation}
The transformed solution thus has the same Newtonian and 
post-Newtonian behavior as the original metric and in addition an 
electromagnetic field. The magnetic field is of order $\Omega^{3}$ 
as $b'$.

Since the mass is of order $\lambda$ in the Newtonian limit, it is 
possible to have an extreme limit here with $M'=Q'$ as in the 
Reissner-Nordstr\"om solution. If we put $1-q^{2}=\kappa \lambda$ 
with $\kappa>0$, 
we get in the limit $\lambda\to0$ 
for (\ref{f'}) and (\ref{Phi'})
\begin{equation}
    f'=\frac{\kappa^{2}}{(\kappa-U_{N})^{2}},\quad \Phi'=
    \frac{U_{N}}{\kappa
    -U_{N}}
    \label{extreme2},
\end{equation}
a static solution similar to the extreme Reissner-Nordstr\"om 
solution, but with a jump in the normal derivatives of the metric 
functions at the disk and non-vanishing $f'$ at the 
origin.  
Since $U_{N}<0$ in the whole spacetime, the solution is regular in the 
exterior of the disk. Thus one gets a non-singular limit in the 
exterior of the disk for $q\to1$ 
in this case.

In the static limit one has $b'=\Im \Phi'=0$, since both are 
proportional to $b$,
\begin{equation}
    f'=\frac{(1-q^{2})f}{(1-q^{2}f)^{2}},\quad \Phi'=
    -q\frac{1-f}{1-q^{2}f}
    \label{static2}.
\end{equation}
The 
Harrison-transformed static solution is thus 
again static with vanishing 
magnetic field but non-zero electric field.

\section{Energy-momentum tensor}
In this section we will study the energy-momentum tensor at 
the disk which is a surface layer. For 
the `cut and glue' techniques, which lead to infinite disks, the 
energy-momentum tensor can be determined using Israel's invariant 
junction conditions. 
In the Einstein-Maxwell 
case we consider the Harrison-transformed counter-rotating disks as 
an example and check the energy conditions. Following \cite{zofka}, 
the matter is interpreted as 
streams of counter-rotating electro-dust where possible.

\subsection{Disk and energy-momentum tensor}
To treat relativistic dust disk, it 
seems best to use Israel's invariant 
junction conditions for matching spacetimes across non-null hypersurfaces 
\cite{israel}. The disk is placed in the equatorial plane and 
the regions $V^{\pm}$ ($\pm \zeta>0$) are matched at the 
hyperplane $\zeta=0$. 
This is possible in Weyl coordinates  since we 
are only considering dust i.e.\ vanishing radial
stresses in the disk. The spacelike unit normal vector of this 
hypersurface in $V^{+}$ is $(n_{\alpha})=e^{k-U}(0,0,1,0)$. The 
extrinsic curvature $K_{AB}$ of this plane in $V^{+}$ is defined as 
$K_{AB}=n_{A||B}$; here capital indices take the values 
0, 1, 3 corresponding to the coordinates $t$, $\rho$, $\phi$, $||$ 
denotes the covariant derivative with respect to 
$s_{AB}$, the metric on the hypersurface. 
According to Israel \cite{israel} the jump $\gamma_{AB}=
K^+_{AB}-K^-_{AB}$ 
in the extrinsic curvature $K_{AB}$
of the hypersurface $\zeta=0$ with respect to its embeddings into 
$V^{\pm}=\{\pm\zeta>0\}$ is related 
to the energy momentum tensor $S_{AB}$ of 
the disk via
\begin{equation}
	-8\pi S_{AB}=\gamma_{AB}-s_{AB}
	\gamma_{C}^{C}
	\label{vac16.1}.
\end{equation}
As a consequence of the field 
equations the energy momentum tensor is divergence free, 
$S^{AB}{}_{||B}=0$. 

The relations (\ref{vac16.1}) lead to 
\begin{eqnarray}
	-4\pi e^{(k-U)}S_{00} & = & \left(k_{\zeta}-2U_{\zeta}
	\right)e^{2U},
	\nonumber \\
    -4\pi e^{(k-U)}( S_{03}-aS_{00}) & = & -\frac{1}{2}a_{\zeta}e^{2U}
	\nonumber ,\\
	-4\pi e^{(k-U)} (S_{33}-2a S_{03}+a^2 S_{00})& = & -k_{\zeta}\rho^2e^{-2U}
	\label{16.7}.
\end{eqnarray}
With these formulas it is straight forward to calculate the 
energy-momentum tensor for a given spacetime. The discontinuity of 
the normal derivatives in the equatorial plane due to the `cut and 
glue' techniques lead as in the Newtonian case to a disk like surface 
layer. There are purely azimuthal tensions in this case. 

Whenever an energy-momentum tensor is worked out in the above way by 
entering with a metric into the Einstein tensor, the question has to 
be addressed whether the matter is physically acceptable. The usual 
criterion one has to check are the energy conditions, see 
\cite{hawkell}.  The weak energy condition implies that the energy 
density must be positive for all observers, $S_{AB}V^{A}V^{B}>0$ 
where $V_{A}$ is an arbitrary timelike vector. The dominant energy 
condition is satisfied if the weak energy condition holds and if 
in addition the flux of energy is positive for any observer, i.e.\ 
that $S^{AB}V_{A}$ is a non-spacelike vector for an  arbitrary timelike 
vector $V_{A}$. An energy-momentum tensor satisfying the dominant energy 
condition will be called physically acceptable, otherwise the matter 
is exotic (negative energy densities, superluminal velocities, \ldots), 
as Bondi called it, 
`not the cheapest material to be bought in the shops'. 

A convenient way to describe the matter of the disk 
in the pure vacuum case was introduced by Bi\v{c}\'ak and Ledvinka in 
\cite{ledvinka}: 
An energy-momentum tensor  $S^{AB}$ with three independent 
components can always be written as 
\begin{equation}
S^{AB}=\sigma_{p}^{*}V^{A}V^{B}+p_{p}^{*}W^{A}W^{B}
\label{2.31a},
\end{equation}
where $V$ and $W$ are the unit timelike respectively spacelike vectors
$(V^{A})=N_{1}(1,0,\omega_{\phi})$ and where 
$(W^{A})=N_{2}(\kappa,0,1)$.  This corresponds to the introduction 
of observers (called $\phi$-isotropic observers (FIOs) in
\cite{ledvinka}) for which the energy-momentum tensor is diagonal. 
The condition $W_{A}V^{A}=0$
determines $\kappa$ and $\omega_{\phi}$ in terms of the metric,
\begin{equation}
    \omega_{\phi}=\frac{g_{33}s_{00}-g_{00}s_{33}+
    \sqrt{(g_{33}s_{00}-g_{00}s_{33})^{2}+4(g_{03}s_{00}-g_{00}s_{03})
    (g_{03}s_{33}-g_{33}s_{03})}}{2(g_{03}s_{33}-g_{33}s_{03})}
    \label{omegaf}
\end{equation}
and 
\begin{equation}
    \kappa= 
    -\frac{g_{03}+\omega_{\phi}g_{33}}{g_{00}+\omega_{\phi}g_{03}}
    \label{kappa}.
\end{equation}
In non-static cases the FIOs rotate with respect to the 
locally non-rotating observers for whom the metric is diagonal. The 
latter rotate with 
the angular velocity $\omega_{l}$ with respect to infinity
\begin{equation}
    \omega_{l}=-\frac{g_{03}}{g_{33}}=\frac{ae^{4U}}{\rho^{2}-
    a^{2}e^{4U}}.
    \label{omegal}
\end{equation}
This quantity is a measure for the dragging of the inertial frames 
with respect to infinity due to the rotating matter in the disk.

If the dominant energy-condition holds and if the pressure is 
positive, one has
$p_{p}^{*}/\sigma_{p}^{*}<1$, and the matter in the disk can be 
interpreted as in \cite{morgan} either as having a purely azimuthal 
pressure or as being made up of two counter-rotating streams of 
pressureless matter with proper surface energy density $\sigma_{p}^{*}/2$
which are counter-rotating with the same angular velocity $
\sqrt{p_{p}^{*}/\sigma_{p}^{*}}$ (which is below 1, the velocity of 
the light),
\begin{equation}
	S^{AB}=\frac{1}{2}\sigma^{*}(U_{+}^{A}U_{+}^{B}+
	U_{-}^{A}U_{-}^{B})
	\label{2.31b}
\end{equation}
where $(U_{\pm}^{A})=U^{*}(v^{A}\pm 
\sqrt{p^{*}_{p}/\sigma^{*}_{p}}w^{A})$ 
is a unit timelike 
vector. We will always adopt the latter interpretation if the 
condition $p_{p}^{*}/\sigma_{p}^{*}<1$ is satisfied which is the 
case in the example \cite{prl2}. 
The energy-momentum tensor (\ref{2.31b}) is just the sum of two 
energy-momentum tensors for dust. Furthermore it can be shown that
the vectors $U_{\pm}$ are geodesic vectors with respect to the inner geometry of 
the disk: this is a consequence of the equation $S^{AB}{}_{||B}=0$ together 
with the fact that $U_{\pm}$ is a linear combination of the Killing vectors. 

Using the `cut and glue' 
formalism, disk sources for all known stationary axisymmetric 
metrics can be given. Sources for static spacetimes were discussed 
in \cite{blk93}, the Kerr spacetime was considered in \cite{ledvinka}. 
If the strip cut-off the spacetime includes the horizon completely 
in the latter 
case, the matter in the disk can be interpreted as dust.  

In the presence of electromagnetic fields,
a discontinuous electromagnetic tensor $F^{\alpha\beta}$ leads to a 
current density $J^{\alpha}$ 
via the Maxwell equations 
\begin{equation}
    F^{A\beta}{}_{;\beta}=\frac{1}{\sqrt{-g}}\left(
    \sqrt{-g}F^{A\beta}\right)_{,\beta}
    =-4\pi J^{A}
    \label{gegen6}.
\end{equation}
Contributions to $J^{A}$ arise only 
via the normal derivatives at the disk. We define the current density 
$j^{A}$ in the 
disk as $s^{AB}$
by the relation $J^{A}=:e^{-2(k-U)}j^{A}\delta(\zeta)$.
The electromagnetic energy-momentum 
tensor does not produce a $\delta$-type contribution to the 
Einstein-Maxwell equations since $F^{\alpha\beta}$ is bounded at the disk. 
With (\ref{gegen6}) we get using  equatorial symmetry (the derivatives 
are taken at $\zeta=0^{+}$)
\begin{equation}
    2\pi j_{0}=-(\Re \Phi)_{\zeta},\quad 2\pi (j_{3}-aj_{0})
    =-\frac{\rho}{f}(
    \Im \Phi)_{\rho}
    \label{gegen7a}.
\end{equation}

The continuity equation (the Bianchi identity)
at the disk $T^{\mu\nu}{}_{;\nu}=F^{\mu\nu}
J_{\nu}=0$ leads to the condition 
\begin{equation}
	g_{00,\rho}s^{00}+2g_{03,\rho}s^{03}+g_{33,\rho}s^{33}=2(
	F_{10}j^{0}+F_{13}j^{3}).
	\label{vac20}
\end{equation}
We remark that 
one can substitute one of the equations (\ref{16.7}) by (\ref{vac20})
in the 
same way as one replaces one of the field equations by the covariant 
conservation of the energy momentum tensor in the case of 
three-dimensional perfect fluids. This makes it possible to eliminate 
$k_{\zeta}$ from (\ref{16.7}) and to treat the energy-momentum tensor at 
the disk purely on the level of the Ernst equation. It is 
straight forward to check 
the consistency of this approach with the help of (\ref{k}). Thus one 
can solve boundary value problems of dust disks without using $k$.

The interpretation of the energy-momentum tensor is not changed with 
respect to the pure vacuum case. However, free particles will no 
longer move on geodesics but on so-called electro-geodesics where the 
Lorentz force is added to the geodesic equation which leads to 
(\ref{vac20}) for each component of dust.
It follows that the matter streams will only
move on electro-geodesics in the FIO frame if
$j_{A}w^{A}=0$, i.e.\ if there are no currents in this frame.
This is in general not the case which implies that the 
FIOs cannot interpret the matter in the disk as freely moving charged 
particles even if the energy conditions are satisfied which was always 
possible in the pure vacuum case. However the matter can still be 
interpreted as a fluid with a purely azimuthal pressure.

Since the splitting of the matter into two streams of dust is not 
unique, an alternative approach is to 
interpret the matter in the disk as two streams of 
particles moving on electro-geodesics in the asymptotically 
non-rotating frame. To this end we make the ansatz \cite{zofka}
\begin{equation}
    s^{AB}=\sigma_{m}^{+}U^{A}_{+}U^{B}_{+}+
    \sigma_{m}^{-}U^{A}_{-}U^{B}_{-},\quad
    j^{A}=\sigma_{e}^{+}U^{A}_{+}+
    \sigma_{e}^{-}U^{A}_{-}
    \label{fio2}
\end{equation}
with $(U^{A}_{\pm})=N_{\pm}
(1,0,\omega_{\pm})$. The angular velocity 
follows from the electro-geodesic equation for each component,
\begin{equation}
    \frac{1}{2}\sigma_{m}^{\pm}(g_{00,\rho}+2g_{03,\rho}\omega_{\pm}+
    g_{33,\rho}\omega_{\pm}^{2})=
    \sigma_{e}^{\pm}(A_{0,\rho}+A_{3,\rho}\omega_{\pm}).
    \label{fio3}
\end{equation} 
Such an interpretation is possible if the angular velocities are 
real. Moreover the velocities $\omega_{\pm}\rho$
in the disk should be smaller than 1 to avoid superluminal velocities,
and the energy densities have to be positive, 
which means we are not interested in tachyonic or otherwise exotic
matter. Relation (\ref{fio2}) leads to 
\begin{equation}
    \sigma_{e}^{+}N_{+}^{2}
    =\frac{j^{3}-\omega_{-}j^{0}}{\omega_{+}-\omega_{-}},
    \quad \sigma_{e}^{-}N_{-}^{2}=
    \frac{j^{3}-\omega_{+}j^{0}}{\omega_{-}-\omega_{+}}
    \label{fio4},
\end{equation}
\begin{equation}
    \sigma_{m}^{+}N_{+}^{2}=\frac{s^{03}-\omega_{-}s^{00}}{
    \omega_{+}-\omega_{-}},\quad
    \sigma_{m}^{-}N_{-}^{2}=\frac{s^{03}-\omega_{+}s^{00}}{
    \omega_{-}-\omega_{+}}
    \label{fio5},
\end{equation}
and
\begin{equation}
    \omega_{-}=\frac{s^{33}-\omega_{+}s^{03}}{s^{03}-\omega_{+}s^{00}}
    \label{fio6}.
\end{equation}
If we enter (\ref{fio3}) with this, we obtain
\begin{equation}
    \omega_{\pm}=\frac{-T_{2}\pm\sqrt{T_{2}^{2}-T_{1}T_{3}}}{T_{1}}
    \label{fio7}
\end{equation}
with 
\begin{eqnarray}
    T_{1} & = & g_{33,\rho}-2A_{3,\rho}\frac{j^{0}s^{03}-j^{3}s^{00}}{
    s^{03}s^{03}-s^{00}s^{33}}
    \nonumber  \\
    T_{2} & = & g_{03,\rho}-A_{0,\rho}\frac{j^{0}s^{03}-j^{3}s^{00}}{
    s^{03}s^{03}-s^{00}s^{33}}-A_{3,\rho}\frac{j^{3}s^{03}-j^{0}s^{33}}{
    s^{03}s^{03}-s^{00}s^{33}}
    \nonumber  \\
    T_{3} & = & g_{00,\rho}-2A_{0,\rho}\frac{j^{3}s^{03}-j^{0}s^{33}}{
    s^{03}s^{03}-s^{00}s^{33}}
    \label{fio8}.
\end{eqnarray}
The densities then follow from (\ref{fio2}) where the continuity 
equation (\ref{vac20}) guarantees that this system can be 
solved. 

Using the above formalism, disk sources for conformastationary 
spacetimes \cite{kbl99}, magnetostatic metrics 
\cite{letelier} and the 
Kerr-Newman family \cite{zofka} have been constructed and discussed.

\subsection{Energy-momentum tensor of the
Harrison-transformed counter-rotating dust disk}
As an example we will now discuss the Harrison-transformed 
counter-rotating dust disk.
It remains to be checked whether and in which range of the 
parameters the energy-momentum tensor 
(\ref{tensor}) is physically acceptable. The above
discussion of the metric 
indicates the extreme behavior of the metric functions for $q$ close 
to one. It is plausible that the matter in the disk which is in the 
present example the
source of such an extreme metric  will in general not be  
physically acceptable. There can be maximal $q$ 
smaller than 1 for given $\lambda$ and $\delta$ which limits the 
physical range of the parameters. 

\begin{figure}[htb]
    \centering
     \epsfig{file=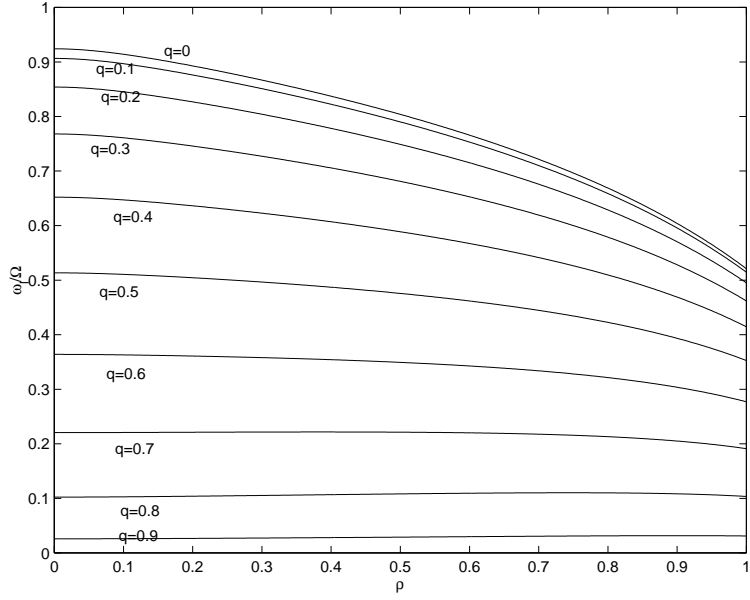,width=10cm}
    \caption{Angular velocity of the locally non-rotating observers 
    with respect to infinity for $\epsilon=0.85$, 
    $\gamma=0.95$, and $q=0,\ldots, 0.9$.}
    \label{ol}
\end{figure}

To discuss the energy-momentum tensor and the currents in the disk, 
it is helpful to use the algebraic relations (\ref{2.9}) to (\ref{x0})
between the metric functions which exist at the disk, and which imply 
similar relations between the transformed potentials. With 
(\ref{12b}) and (\ref{21b}), we get for $\mathcal{S}_{21}$
\begin{equation}
    \mathcal{S}_{21}=\mathcal{E}\bar{\mathcal{E}}S_{12}+ix_{0}f
    \label{21d},
\end{equation}
and thus with (\ref{a'}) for the metric function $a'$
\begin{equation}
    (1-q^{2})^{2}(a-a_{0})'=(a-a_{0})
    \left(1+q^{4}\frac{2}{\delta}\left(\frac{1}{\delta\lambda}
    \left(\frac{1/\lambda^{2}+
    \delta}{\sqrt{1/\lambda^{2}+\delta \rho^{2}}}-\frac{1}{\lambda}
    \right)+\alpha+\frac{\rho^{2}}{2}\right)\right)
    \label{Z'3},
\end{equation}
where $a_{0}'$ is given by (\ref{Z0}). With this function we can can 
calculate the angular velocity of the locally non-rotating observers 
$\omega_{l}$ (\ref{omegal}). 
The dependence of $\omega_{l}$ on $\epsilon$ and $\gamma$ has been 
discussed in \cite{prd4}. As a function of $q$ it is monotonically 
decreasing as can be seen in Fig.~\ref{ol}. The reason for this 
behavior is that the function $f'$ tends to zero in the limit 
$q\to 1$ for finite $\rho$, $\zeta$ whereas $g_{03}'$ changes 
shape but remains finite. Thus the overall behavior of $\omega_{l}$ 
is dominated by $f'$. The deformation of the function $g_{03}'$ 
via $q$ has, however, the consequence that $\omega_{l}$ has its maximum for 
large $q$ no longer at the center at the disk but near the rim.

The energy-momentum tensor at the Harrison-transformed disk can be 
calculated via (\ref{16.7}). Expressing the right-hand sides with the 
help of (\ref{f'}) and (\ref{b'}) via the original functions, we get
\begin{eqnarray}
    s_{1}' & = & 
    \frac{(1-q^{2})^{2}}{N^{2}}\left(\frac{\rho N}{2Z}b_{\rho}
    -(1-q^{4}f^{2}+q^{4}b^{2})f_{\zeta}+2q^{4}fbb_{\zeta}\right),
    \nonumber  \\
    s_{2}' & = & \frac{\rho}{2fN}\left(
    (1-q^{4}f^{2}+q^{4}b^{2})b_{\rho}+2q^{4} 
    bff_{\rho}\right),
    \nonumber  \\
    s_{3}' & = & -\frac{\rho^{3}N}{2(1-q^{2})^{2}f^{2}Z}b_{\rho}
    \label{tensor},
\end{eqnarray} 
where $N=(1-q^{2}f)^{2}+q^{4}b^{2}$.

This implies
\begin{eqnarray}
    2\pi j_{0} & = & \frac{q(1-q^{2})}{N^{2}}\left(-((1-q^{2}f)^{2}-
    q^{4}b^{2})f_{\zeta}+2q^{2}b(1-q^{2}f)b_{\zeta}\right)
    \nonumber  \\
    2\pi (j_{3}-aj_{0}) & = & \frac{\rho q}{(1-q^{2})Nf}\left(
    2q^{2}b(1-q^{2}f)f_{\rho}+((1-q^{2}f)^{2}-
    q^{4}b^{2})b_{\rho}\right)
    \label{gegen7b}.
\end{eqnarray}

\begin{figure}[htb]
    \centering
     \epsfig{file=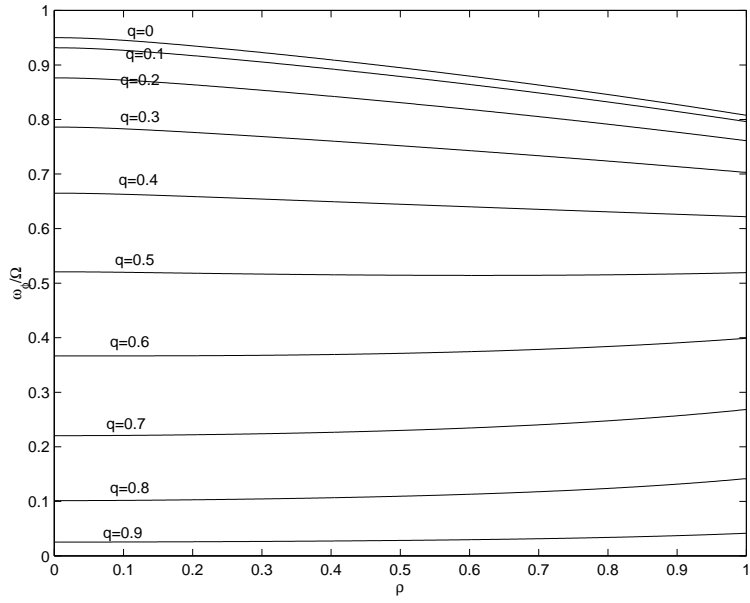,width=10cm}
    \caption{Angular velocity of the FIOs
    with respect to infinity for $\epsilon=0.85$, 
    $\gamma=0.95$, and $q=0,\ldots, 0.9$.}
    \label{of}
\end{figure}

We will interpret the matter in the disk as in \cite{zofka}
in two ways: The FIOs for whom the tensor $s^{AB}$ is diagonal rotate 
with angular velocity $\omega_{\phi}$ with respect to infinity. 
We show $\omega_{\phi}$ for several values of $q$ in Fig.~\ref{of}. It 
can be seen that $\omega_{\phi}$ decreases monotonically with $q$. For 
large $q$ the maximum of the angular velocity is near or at the rim of 
the disk in this example.

\begin{figure}[htb]
    \centering
     \epsfig{file=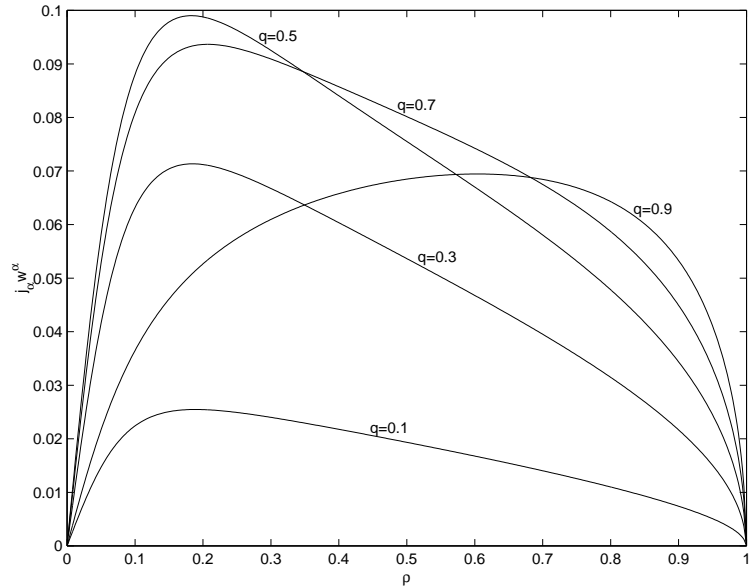,width=10cm}
    \caption{Currents in the frame of the FIOs for $\epsilon=0.85$, 
    $\gamma=0.95$, and $q=0,\ldots, 0.9$.}
    \label{j3}
\end{figure}

Since the energy conditions are satisfied here, 
the FIOs can interpret 
the matter in the disk as a fluid with 
a purely azimuthal pressure. Alternatively they can interpret it as 
being made up of two streams of counter-rotating pressureless matter, 
where the velocity $\Omega_{c}\rho$ in the streams is 
below the speed of light. 
However it can be seen in Fig.~\ref{j3} that there are 
in general currents in the frame of the FIOs. Thus an interpretation 
of the matter by the FIOs as freely moving charged particles is 
not possible.

\begin{figure}[htb]
    \centering
     \epsfig{file=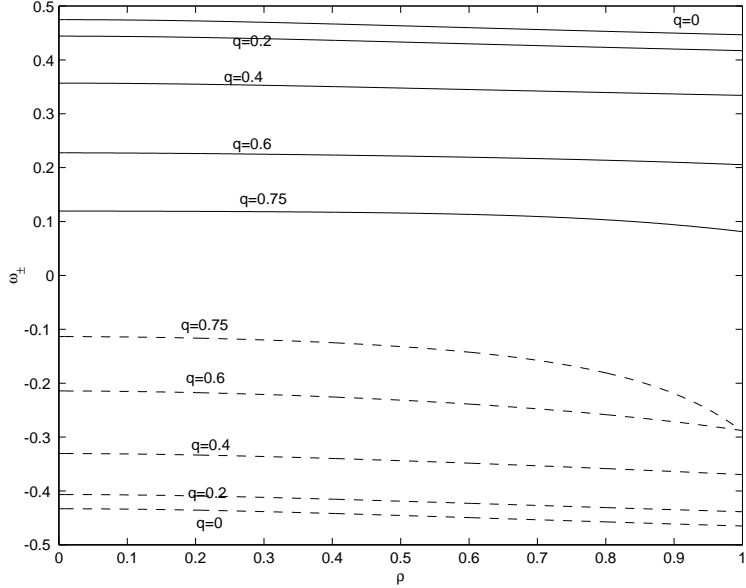,width=10cm}
    \caption{Angular velocities $\omega_{\pm}$ for $\epsilon=0.36$, 
    $\gamma=0.08$, and $q=0,0.2,\ldots,0.8$.}
    \label{omp}
\end{figure}

\begin{figure}[htb]
    \centering
     \epsfig{file=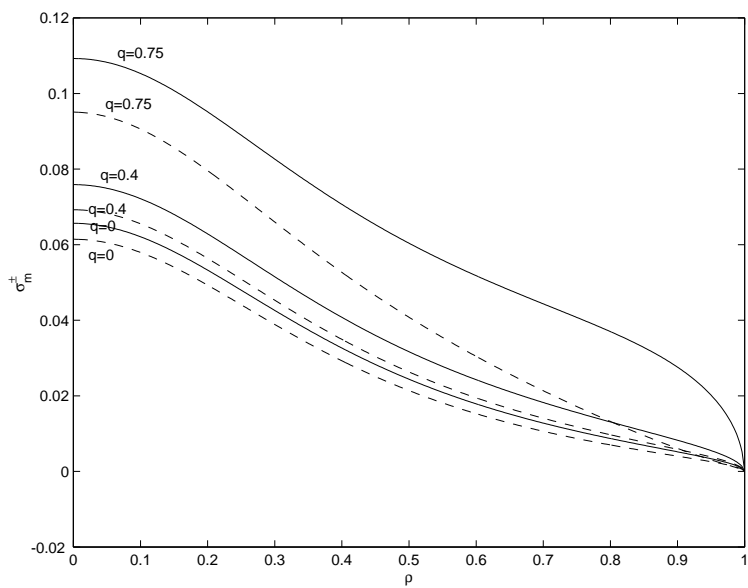,width=10cm}
    \caption{Matter densities  $\sigma_{m}^{+}$ and $\sigma_{m}^{-}$ 
    (dashed) for $\epsilon=0.36$, 
    $\gamma=0.08$, and $q=0,0.4,0.8$.}
    \label{smpm}
\end{figure}

\begin{figure}[htb]
    \centering
     \epsfig{file=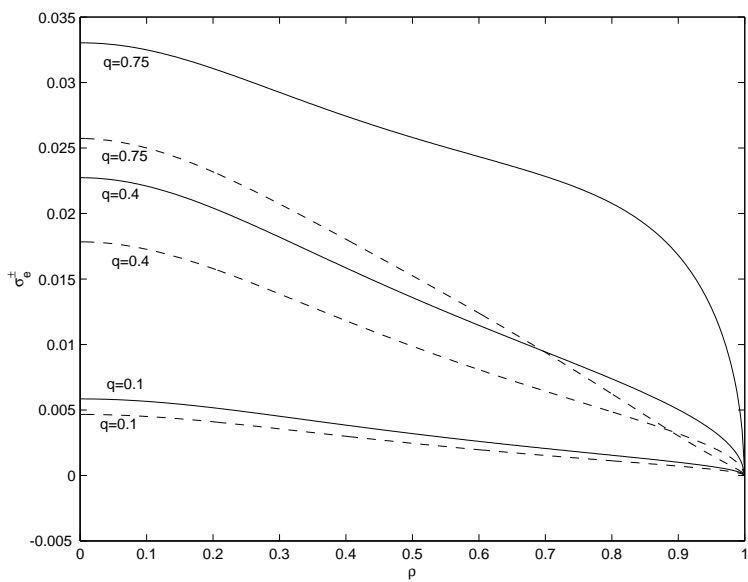,width=10cm}
    \caption{Charge densities  $\sigma_{e}^{+}$ and $\sigma_{e}^{-}$ 
    (dashed) for $\epsilon=0.36$, 
    $\gamma=0.08$, and $q=0,0.4,0.8$.}
    \label{sepm}
\end{figure}

The second interpretation in terms of two streams of electro-dust is 
limited by the conditions that the angular velocities have to be real 
and that the energy densities have to be positive. 
Numerically one finds in the present example
that the angular velocities are real, but in a 
wide range of the parameters there are negative energy densities and 
tachyonic behavior. Already in 
the uncharged case there are infinite velocities in strongly 
relativistic settings with negligible counter-rotation which are due 
to extrema of the metric function $g_{33}$ in the disk. In this case 
the quantity $T_{1}$ in (\ref{fio8}) is zero which leads to a 
diverging $\omega_{-}$. Increasing $q$ only enhances this effect. The 
result is that an interpretation as non-tachyonic counter-rotating 
matter on electro-geodesics with positive energy densities 
is only possible if $q$, $\epsilon$ and 
$\gamma$ are not too large. In other words large values of $q$ are in 
this setting only possible in post-Newtonian or nearly static 
situations. We show plots of the angular velocities $\omega_{\pm}$ 
in Fig.~\ref{omp}
for $\epsilon=0.36$ and $\gamma=.08$ where values of $q$ up to $0.75$ 
are possible.
The corresponding densities are given in Fig. \ref{smpm},
and the charge densities in Fig. \ref{sepm}.
The densities vanish always at the rim of the disk, the 
charge densities are identically zero for $q=0$. For larger values of 
$q$, the angular velocity $\omega_{-}$ becomes bigger and bigger
at the rim of 
the disk and finally diverges. The density $\sigma_{m}^{-}$ is in 
this case negative in the vicinity of the rim of the disk.

\section{Einstein-Maxwell equations and the Riemann-Hilbert problem}
In this section we present recent results on theta functional 
solutions to the Einstein-Maxwell equations which are constructed 
with Riemann-Hilbert techniques. We first obtain the hyperelliptic 
solutions starting from so-called Schlesinger equations. This method 
is then extended to the Einstein-Maxwell case. The keypoint in the 
construction is the solution of the Riemann-Hilbert problem on 
multi-sheeted Riemann surfaces in terms of Szeg\"o kernels by Korotkin 
\cite{koro03}. The 
material in this section is based on \cite{unpublished}. 

\subsection{Riemann-Hilbert problem and Schlesinger equations}
Theorem 3.1 in section 3 offers the possibility to construct matrices 
$\Psi$ with certain analytic properties in the spectral parameter 
which lead to solutions to the Einstein-Maxwell equations. The task is 
to solve a so-called Riemann-Hilbert problem for $\Psi$
to obtain a matrix with prescribed singularities.

The Riemann-Hilbert problem, also known as Hilbert's 21st problem, can 
be stated as follows: Consider a linear Fuchsian ordinary
differential system on $\mathbb{CP}1$ for some $M\times M$ matrix 
$\Psi$,
\begin{equation}
    \frac{d\Psi}{d\gamma}=\sum_{j=1}^{N}\frac{A_{j}}{\gamma-\gamma_{j}}\Psi
    \label{eq:linear},
\end{equation}
where the matrices $A_{j}\in sl(M,\mathbb{C})$ are independent of 
$\gamma$, with  initial condition $\Psi(\gamma=\infty)=I$. 
At the singularities of the equation, the 
solutions will not be single-valued on $\mathbb{CP}1$. 
If one goes on a small loop 
around the singularities, a matrix $\Psi$ which is a fundamental 
solution of the differential system  will change by multiplication 
with some matrix which is called the monodromy matrix. The 
Riemann-Hilbert problem (inverse monodromy problem) 
is the construction of the differential equation for 
a given monodromy matrix, for more details see \cite{bolibrukh}. 

The mathematical formulation of this 
problem for an $M\times M$ matrix reads:
fix an $SL(M,\mathbb{C})$ monodromy representation $\mathcal{M}$ of the fundamental 
group $\pi_1[\mathbb{CP}1\setminus \{\gamma_1,\dots,\gamma_N\}]$.
Choose the standard set of generators $l_1,\dots,l_N$ of 
$\pi_1[\mathbb{CP}1\setminus \{\gamma_1,\dots,\gamma_N\}]$
such that the contour $l_m$ encircles only one singularity $\gamma_{m}$, 
and the relation $l_M\dots l_1=I$
is satisfied. Then the monodromy representation $\mathcal{M}$ is defined 
by the corresponding set of  $N$ $SL(M)$
matrices $\mathcal{M}_1,\dots,\mathcal{M}_N$ satisfying the relation 
$\mathcal{M}_N\dots\mathcal{M}_1=I$.
The Riemann-Hilbert problem is to find a $SL(M,\mathbb{C})$-valued function 
$\Psi(P)$ defined on the universal covering
$X$ of  $\mathbb{CP}1\setminus \{\gamma_1,\dots,\gamma_N\}$ such that 
$\Psi$ gains the right multiplier equal to 
$\mathcal{M}_m$ being analytically continued along contour $l_m$. 

Note that in this formulation the solution 
of the Riemann-Hilbert problem is not unique; different
solutions are related by so-called Schlesinger transformations 
(multiplications from the left with  appropriate
matrix-valued rational functions).
It is known that the Riemann-Hilbert problem cannot be solved 
explicitly in terms of 
known special functions in the generic case. The largest
class of explicitly solvable problems known so far is in the class 
of problems with quasi-permutation monodromies (when each row and 
each column of  monodromy matrix contain exactly one non-vanishing 
entry). The universal cover $X$ is in this case a $M$-sheeted compact 
Riemann surface.

A solution $\Psi\in SL(M,\mathbb{C})$ solves the equation (\ref{eq:linear}). 
The behavior near the singularities is 
given (in general position, i.e.\ 
the difference between the eigenvalues is non-integer 
for each of the matrices $A_{j}$) by the asymptotic expansion of $\Psi$,
\begin{equation}
    \Psi(\gamma)=Q_{j}(I+0(\gamma-\gamma_{j}))(\gamma-\gamma_{j})^{T_{j}} 
    C_{j}
    \label{eq:sing},
\end{equation}
where $Q_{j},C_{j}\in SL(M,\mathbb{C})$ and where $T_{j}$ is a 
diagonal, tracefree matrix. The matrices $M_{j}=C_{j}^{-1}e^{2\pi i 
T_{j}}C_{j}$ are the monodromy matrices.  
The function $\Psi$ is thus only 
single-valued on the universal covering of $\mathbb{C}P1\backslash 
\{\gamma_{1},\ldots,\gamma_{N}\}$. The so-called
isomonodromy condition is 
satisfied if the monodromy matrices are independent of the 
$\gamma_{j}$ which implies that one can change the position of the singularities 
in (\ref{eq:linear}) without affecting the monodromy behavior. In 
general position, this condition implies 
\begin{equation}
    \frac{d\Psi}{d\gamma_{j}}=-\frac{A_{j}}{\gamma-\gamma_{j}}\Psi
    \label{eq:iso}.
\end{equation}
The consistency of these equations with (\ref{eq:linear}) leads to the 
Schlesinger system \cite{schlesinger} for the residues $A_{j}$,
\begin{equation}
    \frac{dA_{j}}{d\gamma_{i}}=\frac{[A_{i},A_{j}]}{\gamma_{i}-\gamma_{j}}, \quad 
    i\neq j, \quad \frac{dA_{i}}{d\gamma_{i}}=-\sum_{j\neq i}^{}
    \frac{[A_{i},A_{j}]}{\gamma_{i}-\gamma_{j}}.
    \label{eq:schlesinger}.
\end{equation}

To solve the Riemann-Hilbert problem for quasi-permutation monodromies 
\cite{koro03}, we 
introduce two further objects on $X$: the prime form 
$E(P,Q)$ has the property that it vanishes at exactly one point on 
$X$, namely when $P=Q$. Since the surface is compact, 
there can be no function with this property (the Riemann vanishing 
theorem states that Riemann theta 
functions have exactly $g$ zeros). The prime form is a
$(-\frac{1}{2},-\frac{1}{2})$ differential on the bundle
$X\times X$
\begin{equation}
    E(a,b) = 
    \frac{\Theta_{*}(\int_{a}^{b})}{h_{\Delta}(a)h_{\Delta}(b)}
    \label{eq:prime}
\end{equation}
where 
$*\equiv [\mathbf{p}^{*}\mathbf{q}^{*}]$ denotes a non-singular odd 
characteristic, and where the spinor $h_{\Delta}$ is given by 
$h^{2}_{\Delta}(a)=\sum_{\alpha=1}^{N}\partial_{z_{\alpha}}
\Theta_{*}(0)d\omega_{\alpha}(\tau_{a})$ ($\tau_{a}$ denotes local 
coordinates in the vicinity of a point $a$). 
In local coordinates the prime form reads
\begin{eqnarray*}
	E(a,b) & = & \frac{\tau_{a}-\tau_{b}}{\sqrt{d\tau_{a}}\sqrt{d\tau_{b}}}
	+\ldots .
\end{eqnarray*}
The Szeg\"o kernel $S_{\mathbf{p}\mathbf{q}}(P,Q)$ is defined as 
\begin{equation}
    S_{\mathbf{p}\mathbf{q}}(P,Q) = 
    \frac{\Theta_{\mathbf{p}\mathbf{q}}
    (\mathbf{z}+\int_{P}^{Q})}{\Theta_{\mathbf{p}\mathbf{q}}
    (\mathbf{z}) E(P,Q)}
    =\frac{\sqrt{d\tau_{P}}\sqrt{d\tau_{Q}}}{\tau_{P}-\tau_{Q}}+\ldots
    \label{eq:szego}.
\end{equation}
As can be seen from the representation in local coordinates, it 
can be viewed as a generalization of 
the Cauchy kernel to Riemann surfaces. 

In terms of the Szeg\"o kernel the solution to the Riemann-Hilbert 
problem was written  by Korotkin \cite{koro03} in the form
\begin{equation}
    \Psi_{jk}(K,\lambda)=S_{\mathbf{p}\mathbf{q}}(K^{(j)},\lambda^{(k)})
    E_{0}(K,\lambda)\quad j,k=1,\ldots,M
    \label{eq:solu},
\end{equation}
where $K^{(j)}$ denotes a point on the jth sheet with projection $K$ 
on $\mathbb{CP}1$, and where 
\begin{equation}
    E_{0}(K,\lambda)=\frac{K-\lambda}{\sqrt{dK}\sqrt{d\lambda}}
    \label{eq:prime0}
\end{equation}
is the prime form on $\mathbb{CP}1$. Obviously $\Psi$ defined in 
(\ref{eq:solu}) is not a single-valued function on $\mathbb{CP}1$ (it 
is only single-valued on $X$) since it is a solution of the 
Riemann-Hilbert problem.

\subsection{Vacuum case}
It is instructive to reconsider first the 
two-dimensional vacuum case. In section 4 we treated
the Ernst equation as the integrability condition for 
the linear system (\ref{a1}) which had the advantage that the Ernst 
potential is given as one component of $\Psi(\infty^{+})$. Several 
linear systems for the Ernst equation are known in the literature 
which are related through gauge transformations, see \cite{cosgrove}. 
In this section we will use a linear system which makes the symmetry 
of the Ernst equation obvious since the 
matrix of the linear system is an element of the coset space
$SL(2,\mathbb{R})/SO(2)$ (in an abuse of 
notation we will call this matrix also $\Psi$),
\begin{equation}
    \Psi_{\xi}=\frac{\mathcal{G}_{\xi}\mathcal{G}^{-1}}{1-\gamma}\Psi
    ,\quad
    \Psi_{\bar{\xi}}=\frac{\mathcal{G}_{\bar{\xi}}
    \mathcal{G}^{-1}}{1+\gamma}\Psi
    \label{eq:linsys2},
\end{equation}
where 
\begin{equation}
    \gamma(K,\xi,\bar{\xi})
    =\frac{2}{\xi-\bar{\xi}}\left(K-\frac{\xi+\bar{\xi}}{2}+
    \sqrt{(K-\xi)(K-\bar{\xi})}\right)
    \label{eq:gamma},
\end{equation}
and where 
\begin{equation}
    \mathcal{G}=\frac{1}{\mathcal{E}+\bar{\mathcal{E}}}\left(
    \begin{array}{cc}
	2 & i(\mathcal{E}-\bar{\mathcal{E}})   \\
	i(\mathcal{E}-\bar{\mathcal{E}}) & 2\mathcal{E}\bar{\mathcal{E}}
    \end{array}
 \right)
 \label{eq:G}.
\end{equation}
It is straight forward to adapt theorem 3.1 to the linear system 
(\ref{eq:linsys2}) and to the 
solution of a Riemann-Hilbert problem as is done in the
following theorem by Korotkin and Nicolai \cite{koronic}:\\
\begin{theorem}\label{thm:2}
    Let the matrices $A_{j}\in sl(2,\mathbb{C})$ satisfy the Schlesinger system 
    (\ref{eq:schlesinger}), and let $\Psi(\gamma)$ be the 
    corresponding solution of (\ref{eq:linear}). Suppose that 
    the matrix $\Psi$ satisfies the additional conditions
    \begin{equation}
	\Psi^{T}(1/\gamma)\Psi^{-1}(0)\Psi(\gamma)=I
	\label{eq:red}
    \end{equation}
    and 
    \begin{equation}
	\Psi(-\bar{\gamma})=\bar{\Psi}(\gamma)
	\label{eq:real}.
    \end{equation}
   Let $\gamma_{j}=\gamma(K_{j},\xi,\bar{\xi})$ with $K_{j}\in 
   \mathbb{C}$ independent of $\xi$, $\bar{\xi}$. Then the matrix 
    \begin{equation}
	\mathcal{G}=\Psi(\gamma=0,\xi,\bar{\xi})
	\label{eq:inf}
    \end{equation}
    is real and symmetric and satisfies the Ernst equation, and the 
    function $\Psi$ satisfies the system (\ref{eq:linsys2}). 
\end{theorem}
The proof is analogous to the one for theorem 3.1, see \cite{koronic}. 
The solution of the Riemann-Hilbert problem gives a $SL(2,\mathbb{C})$ matrix 
$\Psi$, the conditions (\ref{eq:red}) and (\ref{eq:real}) ensure that 
$\Psi$ is in the coset $SL(2,\mathbb{R})/SO(2)$. The involution $\sigma$ 
which interchanges the sheets on $\mathcal{L}$ acts as $\gamma\to 
1/\gamma$, condition (\ref{eq:red}) is thus equivalent to the 
reduction condition III. The reality condition (\ref{eq:real}) was 
previously incorporated in the normalization condition IV.

In the two-dimensional vacuum case, the covering surface $X$ is 
hyperelliptic given by 
$\hat{\mu}^{2}=\prod_{j=1}^{N}(\gamma-\gamma_{j})$ (we assume that 
$N$ is even). Since $\gamma$ depends on the spectral parameter $K$ 
via (\ref{eq:gamma}) which lives on the two-sheeted surface 
$\mathcal{L}$, the surface is the four-sheeted 
covering of the complex plane  $\hat{\mathcal{L}}$ 
with Hurwitz diagram Fig.~\ref{hurwitz} of section 4. The factorization 
$\hat{\mathcal{L}}/\sigma$ leads as before to the hyperelliptic 
surface $\mathcal{L}_{H}$.

The corresponding solution of the Ernst equation can be obtained via 
(\ref{eq:inf}). One obtains 
\begin{equation}
    \mathcal{E}=\frac{\Theta_{\mathbf{p}\mathbf{q}}(
    \omega(\infty^{+}))}{\Theta_{\mathbf{p}\mathbf{q}}(
    \omega(\infty^{-}))}
    \label{eq:ernstpq}.
\end{equation}
The relation to the previous form of the Ernst potential can be 
established as follows: Let $g=\tilde{g}+n$ and 
$p_{\tilde{g}+j}=h_{j}\in \mathbb{R}$, $q_{\tilde{g}+j}=0$ 
for $j=1,\ldots,n$. 
Consider the limit of collapsing branch cuts for $j>\tilde{g}$, i.e.\ 
$E_{\tilde{g}+j}\to F_{\tilde{g}+j}$. In this limit all quantities 
entering (\ref{eq:ernstpq}) can be expressed in terms of quantities
(denoted with a tilde)
on the surface $\tilde{\mathcal{L}}_{H}$ of genus $\tilde{g}$ given 
by $\tilde{\mu}^{2}=(K-\xi)(K-\bar{\xi})
\prod_{i=1}^{\tilde{g}}(K-E_{i})(K-F_{i})$, the surfaces 
$\mathcal{L}_{H}$ with the collapsing cuts removed. The holomorphic 
differentials have the limit, see \cite{fay}, 
\begin{equation}
    d\omega_{i}\to d\tilde{\omega}_{i}, \quad i=1,\ldots,\tilde{g},
    \quad d\omega_{i}\to d\tilde{\omega}_{E_{i}^{-}E_{i}^{+}},\quad 
    i=\tilde{g}+1,\ldots,n
    \label{eq:diffdeg}.
\end{equation}
In other words the holomorphic differentials of $\mathcal{L}_{H}$ 
become holomorphic differentials on $\tilde{\mathcal{L}}_{H}$ and 
differentials of the third kind with poles at the collapsed branch 
cuts. Since the $b$-periods of differentials of the third kind can be 
expressed in terms of the Abel map of the poles, see e.g.\ 
\cite{dubrovin}, 
\begin{equation}
    \int_{b_{j}}^{}d\omega_{PQ}=\omega_{j}(P)-\omega_{j}(Q)
    \label{eq:bperiod},
\end{equation}
one can use formula (\ref{theta2}) to get for (\ref{eq:ernstpq}) 
\begin{equation}
    \mathcal{E}=\frac{\tilde{\Theta}_{\tilde{\mathbf{p}}
    \tilde{\mathbf{q}}}(\tilde{\omega}(\infty^{+})+\sum_{j=1}^{n}
    h_{j}(\omega(E_{j}^{-})-\omega(E_{j}^{+})))}{
    \tilde{\Theta}_{\tilde{\mathbf{p}}
	\tilde{\mathbf{q}}}(\tilde{\omega}(\infty^{-})+\sum_{j=1}^{n}
	h_{j}(\omega(E_{j}^{-})-\omega(E_{j}^{+})))}
	\exp\left(\sum_{j=1}^{n}h_{j}\int_{\infty^{-}}^{\infty^{+}}
	d\omega_{E_{j}^{-}E_{j}^{+}}\right)
    \label{eq:ernstdeg}.
\end{equation}
By taking the limit $\sum_{j=1}^{n}\to \int_{\Gamma}^{}$ from a sum to 
a line integral over the $E_{j}$, 
we get after a partial integration (we assume $\ln G$ 
vanishes as the limits of integration) and the 
identification $h(K)=\partial_{K}(\ln G)$ formula (\ref{ernst2}) 
for (\ref{eq:ernstdeg}) where we have used $\int_{A}^{B}d\omega_{PQ}
=\int_{P}^{Q}d\omega_{AB}$. For details of the above construction  see 
\cite{koromat}.

\subsection{The Einstein-Maxwell case}
The construction of algebro-geometric solutions to the 
Einstein-Maxwell equations can be carried out as above for the vacuum 
case. The associated linear system has the form (\ref{eq:linear}) 
where this time the matrix $\mathcal{G}\in SU(2,1)/SU(2)$. 
The matrix $\Psi$ is constructed as the solution of a 
Riemann-Hilbert problem as before. This time the covering surface is 
three-sheeted as noted in section 3. Since the spectral parameter
$\gamma$ varies on the two-sheeted surface $\mathcal{L}$, the 
covering surface $\hat{\mathcal{L}}$ is now six-sheeted. 
Factorizing with respect to 
the involution $\sigma$ of $\mathcal{L}$ leads to a three-sheeted 
surface $\mathcal{L}_{3}$ 
which replaces the hyperelliptic surface $\mathcal{L}_{H}$ in the 
vacuum case. 
All branch points of this
surface are constant except the two branch points $\xi$ and 
$\bar{\xi}$ (the fixed points of the involution $\sigma$), 
where the first two sheets are glued. 
If the third sheet `detaches' from the first two, one gets the 
vacuum solutions.

The reduction conditions to obtain $\mathcal{G}\in SU(2,1)/SU(2)$ 
imply that $\mathcal{L}_{3}$ is invariant with respect to the 
holomorphic involution $\sigma$, acting on every sheet of $\mathcal{L}_{3}$ as 
$\gamma\to 1/\gamma$. In addition $\mathcal{L}_{3}$ has to be 
invariant with respect to the anti-holomorphic
involution $\tau$, acting as $\gamma\to-\bar{\gamma}$ on the 
third sheet of 
$\mathcal{L}_{3}$; on the first and second sheets $\tau$ acts as 
a superposition  of the 
conjugation $\gamma\to-\bar{\gamma}$  with interchange of the first 
and second sheets. 

The solution of the Riemann-Hilbert problem is again provided by 
(\ref{eq:solu}) for $M=3$.
The Ernst potentials can be obtained from $\Psi(\infty^{(1)})$ as before. 
Details of the construction will be published in a forthcoming paper 
by Korotkin. 
The class of algebro-geometrical solutions of the Einstein-Maxwell
system constructed in \cite{koro88} is a partial case of the
above construction
when the third sheet is attached to the first and second sheets only
via branch points of multiplicity two
(i.e. all three sheets are glued together at these points).

\section{Conclusion and Outlook}
In this article we reviewed exact solutions to the Einstein-Maxwell 
equations with disk sources. The `cut and glue' techniques start with 
known exact solution of the equations with equatorial symmetry. The 
hyperplanes $\zeta_{0}$ and $-\zeta_{0}$ are identified. This 
corresponds to removing a strip from the spacetime. At the newly 
formed equatorial plane, the normal derivatives of the metric 
functions will be discontinuous which leads via the field equations 
to a $\delta$-type energy-momentum tensor. These infinitesimally thin 
disks have infinite extension. For asymptotically flat spacetimes, 
the mass of the disks is however finite. The matter in the disk has 
to satisfy the energy conditions in order to be physically acceptable. 
If this is the case, 
an interpretation of the matter in the disk as two streams of counter-rotating 
dust is possible if the velocities in the disk are subluminal. In the 
absence of electromagnetic fields, the particles of the streams move 
on geodesics. In the non-static Einstein-Maxwell case, the electro-geodesic 
equations leads to additional conditions on the matter which can only 
be satisfied in a certain range of the physical parameters. 
`Cut and glue'-techniques thus provide a generally 
applicable method to find (infinite) 
disk sources to known spacetimes. They do 
not provide, however, a method to construct new solutions to the 
Einstein-Maxwell equations.

The complete integrability of the stationary axisymmetric Einstein-Maxwell 
equations offers powerful techniques to generate new solutions. The 
most efficient methods known up to day arise from algebraic geometry 
and lead to explicit solutions in terms of theta functions associated 
to certain Riemann surfaces. In the pure vacuum case, these surfaces 
are hyperelliptic, which makes the use of the powerful hyperelliptic 
calculus possible. These solutions contain the `solitonic' 
solutions as the Kerr solution as limiting cases. An important feature 
of the algebro-geometric solutions  
is that general regularity theorems can be established. Thus the 
question of global regularity in the exterior of the disk source can 
be addressed. With these techniques, disks of finite extension can be 
constructed as the presented family of counter-rotating dust disks 
which leads to new solutions to the Einstein equations. 
The task is then to solve boundary value problems at the disk arising 
from physical models. 

In the Einstein-Maxwell case, hyperelliptic disk solutions were 
constructed by exploiting the $SU(2,1)$-symmetry of the field equations. 
This leads to charged disks. By studying the asymptotics of the so 
generated spacetime, it was shown that the disks always have a 
non-vanishing total charge and a gyromagnetic ratio of 2. To generate 
solutions without total charge but with 
non-trivial magnetic field, it seems 
necessary to study theta functions on three-sheeted surfaces. We 
gave a review on recent results on the Riemann-Hilbert problem on 
multi-sheeted surfaces which can be applied to the Einstein-Maxwell 
case.

In order to obtain disk solutions on three-sheeted surfaces, the 
solutions to Riemann-Hilbert problems have to be adapted to the 
Einstein-Maxwell equations. The general structure of the physically 
interesting Riemann surfaces has to be explored. An important point 
is to establish the analyticity properties of the solutions. If this 
is achieved, disk solutions can be studied. In order to construct 
solutions to prescribed matter models, one will have to adopt the 
algebraic approach \cite{prd3,tmp2} to the three-sheeted case. For a 
complete understanding of the solution, it will be necessary to extend 
the numerical code for hyperelliptic theta functions \cite{prd4} to 
these surfaces.

\thanks{
I thank D.~Korotkin for many discussions and hints, and M.~King for 
carefully reading the manuscript. This work was 
supported by the Schloessmann foundation.
}

\end{document}